\documentclass[twocolumn,twocolappendix,tighten,times]{aastex7}

\usepackage{comment}
\usepackage{caption}

\submitjournal{ApJL}

\shorttitle{NGC 1052 trail dwarf distances}
\shortauthors{Tang et al.}

\begin{document}

\title{New Measurements of Distances to Galaxies in the NGC 1052 Field with the Hubble and James Webb Space Telescopes: Testing the Bullet-Dwarf Origin of the Trail}

\correspondingauthor{Yimeng Tang}
\email{ymtang@ucsc.edu}

\author[0000-0003-2876-577X]{Yimeng Tang}
\affiliation{Department of Astronomy \& Astrophysics, University of California Santa Cruz, 1156 High Street, Santa Cruz, CA 95064, USA}
\email{ymtang@ucsc.edu}

\author[0000-0002-5259-2314]{Gagandeep S. Anand}
\affiliation{Space Telescope Science Institute, 3700 San Martin Drive, Baltimore, MD 21218, USA}
\email{ganand@stsci.edu}

\author[0000-0003-2473-0369]{Aaron J.\ Romanowsky}
\affiliation{Department of Astronomy \& Astrophysics, University of California Santa Cruz, 1156 High Street, Santa Cruz, CA 95064, USA}
\affiliation{Department of Physics \& Astronomy, San Jos\'e State University, One Washington Square, San Jose, CA 95192, USA}
\email{aaron.romanowsky@sjsu.edu}

\author[0000-0002-8282-9888]{Pieter G.\ van Dokkum}
\affiliation{Department of Astronomy, Yale University, New Haven, CT 06511, USA}
\email{pieter.vandokkum@yale.edu}

\author[0000-0001-9742-3138]{Kevin A.\ Bundy}
\affiliation{Department of Astronomy \& Astrophysics, University of California Santa Cruz, 1156 High Street, Santa Cruz, CA 95064, USA}
\email{kbundy@ucsc.edu}

\begin{abstract}

NGC~1052-DF2 and DF4 are two ultra-diffuse galaxies deficient in dark matter (DM), and reported as part of a remarkable linear trail of dwarf galaxies in the NGC~1052 field. Recently, NGC~1052-DF9 has been identified as the third galaxy missing DM along the trail. This structure may have been formed in a high-velocity head-on collision between two gas-rich dwarfs, known as the ``bullet-dwarf'' scenario. However, the trail overlaps in projection with a foreground system, the NGC~1035 group at $\sim13$~Mpc, raising suspicions that the trail is an artifact of this superposition. DF2 and DF4 have been found to be at distances of $21.7\pm1.2$ and $20.0\pm1.6$ Mpc, respectively, using the tip of the red giant branch (TRGB) method with deep Hubble Space Telescope (HST) imaging, but the distances to other trail dwarfs remain unknown. In this Letter, we use HST imaging to obtain surface brightness fluctuation (SBF) distance estimates for eight candidate trail dwarfs, as well as for the giant galaxies NGC~1052 and NGC~1035. We find that the dwarfs are all at $\sim$20 Mpc, and are not associated with the foreground NGC 1035 group. However, for DF2, we derive an SBF distance of $17.7\pm1.4$ Mpc, inconsistent with the published HST TGRB distance ($21.7\pm1.2$ Mpc). Meanwhile, James Webb Space Telescope (JWST) observations of DF2 offer a second, and potentially more accurate, TRGB distance of $17.6\pm0.6$ Mpc. While this value matches our SBF result, it is clear that uniform JWST imaging of the remaining trail dwarfs is critically needed.

\end{abstract}


\section{Introduction} \label{sec:intro}

The origin of galaxies that appear to lack DM remains one of the most intriguing puzzles in near-field cosmology. The discovery of two ultra-diffuse galaxies (UDGs) NGC 1052-DF2 and -DF4, which exhibit dynamical masses consistent with their stellar masses alone \citep{vanDokkum2018a,vanDokkum2019,Wasserman2018,Danieli2019,Emsellem2019,Keim2022,Shen2023}\footnote{DF2 was discovered by \cite{Fosbury1978} and first cataloged by \cite{Karachentsev2000}.}, poses a challenge to our understanding of the connection between galaxy formation and DM halos within the $\Lambda$CDM cosmology. Both galaxies are also unusual in hosting overluminous globular clusters (GCs; \citealt{vanDokkum2018b,vanDokkum2019,Shen2021a}), and share similar age, metallicity, and single-burst star formation history \citep{vanDokkum2022b,Buzzo2022,Tang2025a}.

All these striking similarities suggest a shared origin for DF2 and DF4. The ``bullet dwarf'' scenario stands out so far as the best model matching all the observations \citep{Silk2019,Shin2020,Lee2021,Otaki2022,Lee2024}. In this scenario, a high-velocity collision between two gas-rich progenitors separates shocked gas from DM, producing diffuse DM-free dwarfs. At the same time, unusually massive star clusters form in highly compressed gas clumps. Crucially, the relative line-of-sight velocities, distances, and stellar ages of DF2 and DF4 are all consistent with the prediction of a collision event \citep{vanDokkum2022a}. Such an unusual pair of galaxies may not be unique, as another pair of dwarf galaxies with similar properties (FCC~224 and FCC~240) has recently been discovered in the Fornax Cluster field, which might also have originated from a bullet dwarf event \citep{Buzzo2025,Buzzo2026,Tang2025b}.

\cite{vanDokkum2022a} went on to find that DF2 and DF4 are part of a remarkable linear trail of dwarf galaxies in the NGC 1052 field, which was predicted in bullet-dwarf simulations \citep{Shin2020} and subsequently reproduced by \cite{Lee2024}. Recent observational evidence supports the physical significance of the trail. \cite{Tang2025a} showed that the trail dwarfs are distinct from other dwarfs in the NGC 1052 group, with significantly older ages and higher metallicities. Also, they exhibit a possible alignment of their photometric major axes parallel to the trail. Furthermore, \cite{Keim2025} confirmed that another five trail dwarfs are kinematically connected to DF2 and DF4, matching the theoretical predictions \citep{Lee2024}. Importantly, DF9 has recently been identified as the third galaxy missing dark matter on the trail \citep{Keim2026}, hosting a nuclear star cluster but lacking an unusual GC system as seen in DF2 and DF4.

Given the emerging picture of a novel dwarf formation mechanism, measuring the line-of-sight distances to these trail dwarfs remains crucial to confirming conclusively the existence of a linear structure in three-dimensional space -- specifically, whether or not their line-of-sight distances correlate strongly with their projected location on the trail \citep{Lee2024}. Also, since the relative line-of-sight distance of $1.7 \pm 0.5$ Mpc between DF2 and DF4 \citep{Shen2021b,Shen2023} is larger than the virial diameter of NGC~1052 group ($\sim 0.8$~Mpc; \citealt{Forbes2019}), most of the trail dwarfs would not be in the NGC~1052 group in the bullet dwarf scenario. Alternatively, the trail dwarfs could simply be random members in the NGC 1052 group that have coincidentally aligned in projection, despite the very low probability \citep{vanDokkum2022a}. It has also been proposed that some of these dwarfs might be part of a foreground group associated with NGC~1035 at $\sim 13$ Mpc, and thus the trail could be an artifact of two galaxy groups overlapping in projection \citep{Monelli2019,Montes2020,Golini2024}.

The line-of-sight distances of DF2 and DF4 have been measured using the tip of the red giant branch (TRGB) method with deep Hubble Space Telescope (HST) imaging ($21.7 \pm 1.2$ and $20.0 \pm 1.6$ Mpc; \citealt{Danieli2020,Shen2021b,Shen2023}). However, the redshift-independent distances to the other trail members remain unknown. Fortunately, HST observations are now available for these dwarfs: although the imaging is too shallow for the TRGB approach, it is well-suited for surface brightness fluctuation (SBF) measurements. SBF is an extragalactic distance indicator that measures the pixel-to-pixel surface brightness variance arising from unresolved stellar populations, which decreases with distance. \cite{Cohen2018} previously conducted SBF analysis for some of the trail dwarfs with HST imaging. In this Letter, we apply the SBF method to an expanded sample of HST images to derive distances for the trail dwarfs. While SBF-based distances have much larger uncertainties than those of TRGB, they are sufficient for a definitive test of the overlapping group scenario, and could provide further clues to the nature of the trail.

Additionally, deep archival data from the James Webb Space Telescope (JWST) allow us to obtain another independent, high-precision TRGB distance for DF2. This measurement is expected to serve as a critical benchmark for assessing systematic uncertainties in other distance indicators and lays the groundwork for precisely locating more trail dwarfs in the future.

The rest of this letter is structured as follows. In Section \ref{sec:sbf}, we describe our distance measurements for the trail dwarfs based on the SBF method using HST imaging. In Section \ref{sec:trgb}, we present the TRGB distance measurement for DF2 from JWST observations. The results are shown in Section \ref{sec:results}, followed by discussion and summary in Section \ref{sec:discussion}. All magnitudes in this work are in the AB system, except for the JWST magnitudes in Section \ref{sec:trgb}, which are given in the Vega system.

\begin{figure*}
\centering
\includegraphics[width=1.0\textwidth]{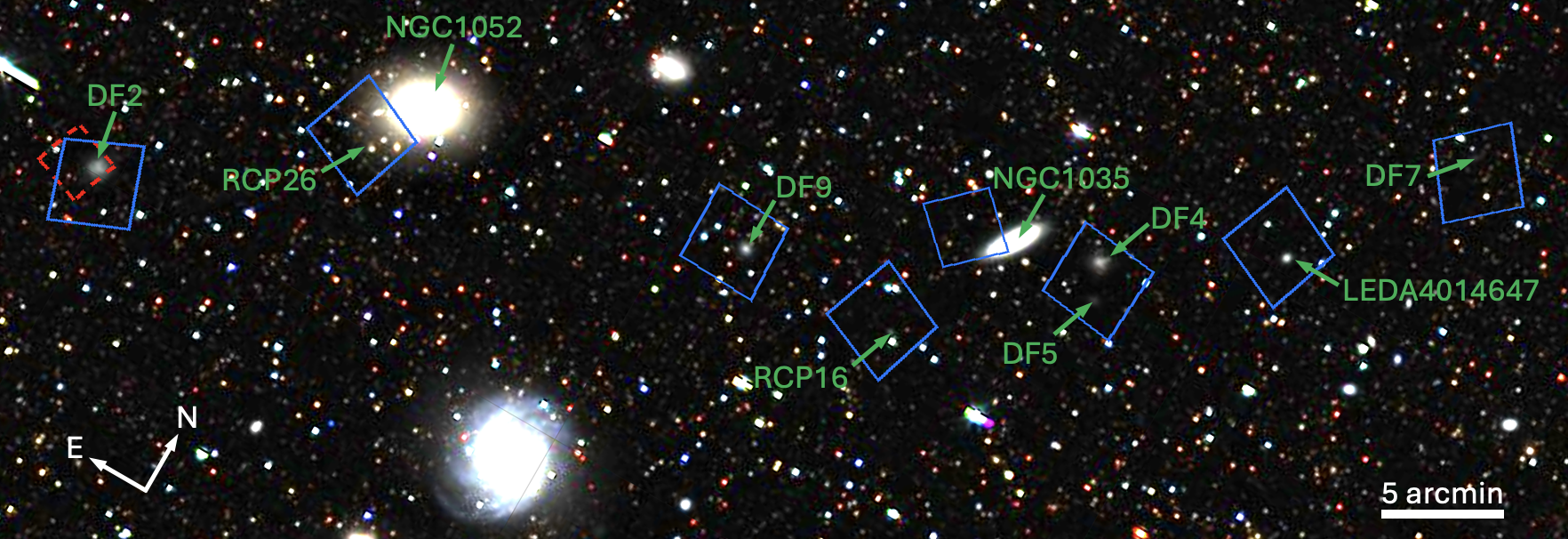}
\includegraphics[width=1.0\textwidth]{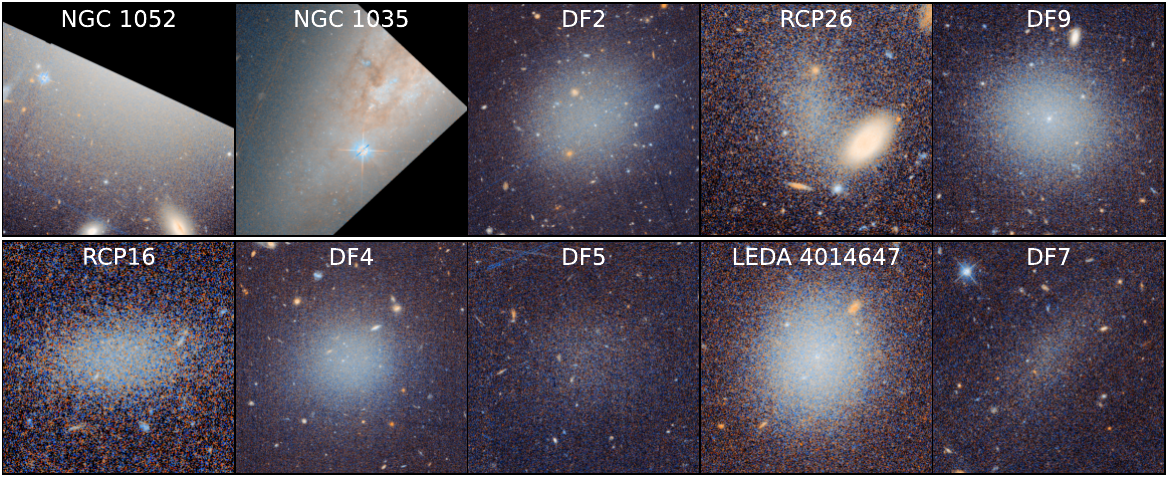}
\caption{{\bf Top panel:} DECaLS imaging of the NGC~1052 field, with NGC~1052, NGC~1035, and eight of the trail dwarfs marked. The HST footprints are overplotted as blue boxes, while the JWST footprint for DF2 is shown as the red dashed box. {\bf Bottom two rows:} The image gallery of NGC~1052, NGC~1035, and eight trail dwarfs studied in the work (not to scale). The pseudocolor images are created using the HST F606W and F814W bands.}
\label{fig:ngc1052_field}
\end{figure*}

\section{Surface Brightness Fluctuation Distances} \label{sec:sbf}

\begin{figure*}
\centering
\includegraphics[width=0.9\textwidth]{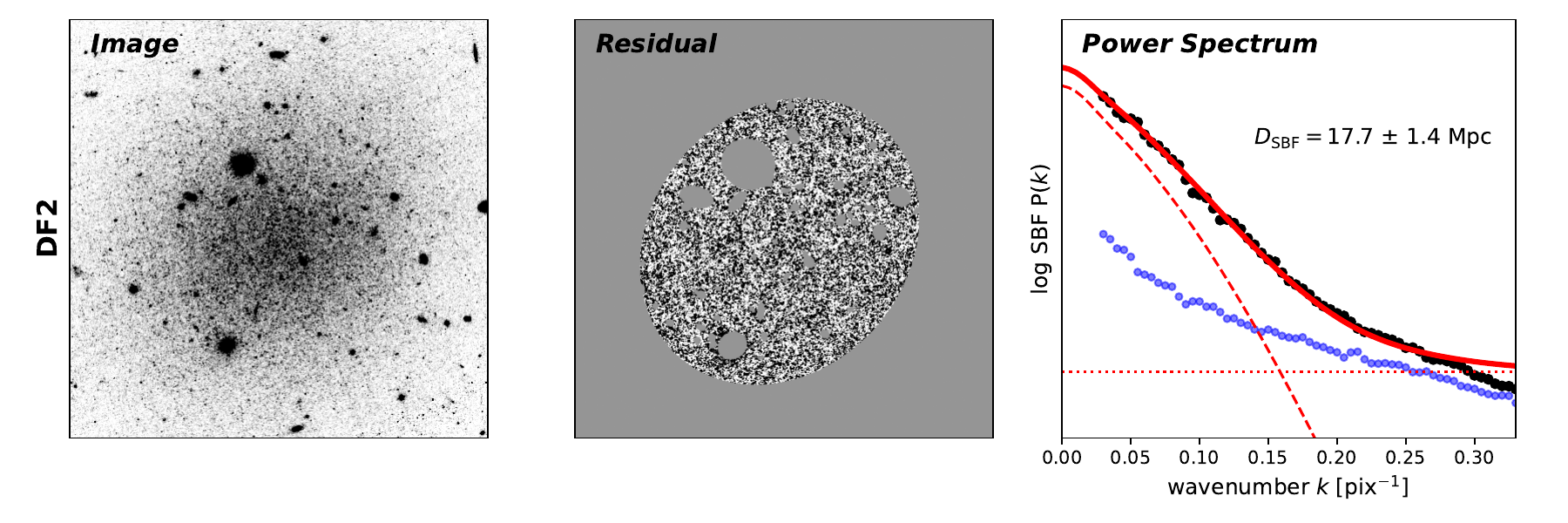}
\includegraphics[width=0.9\textwidth]{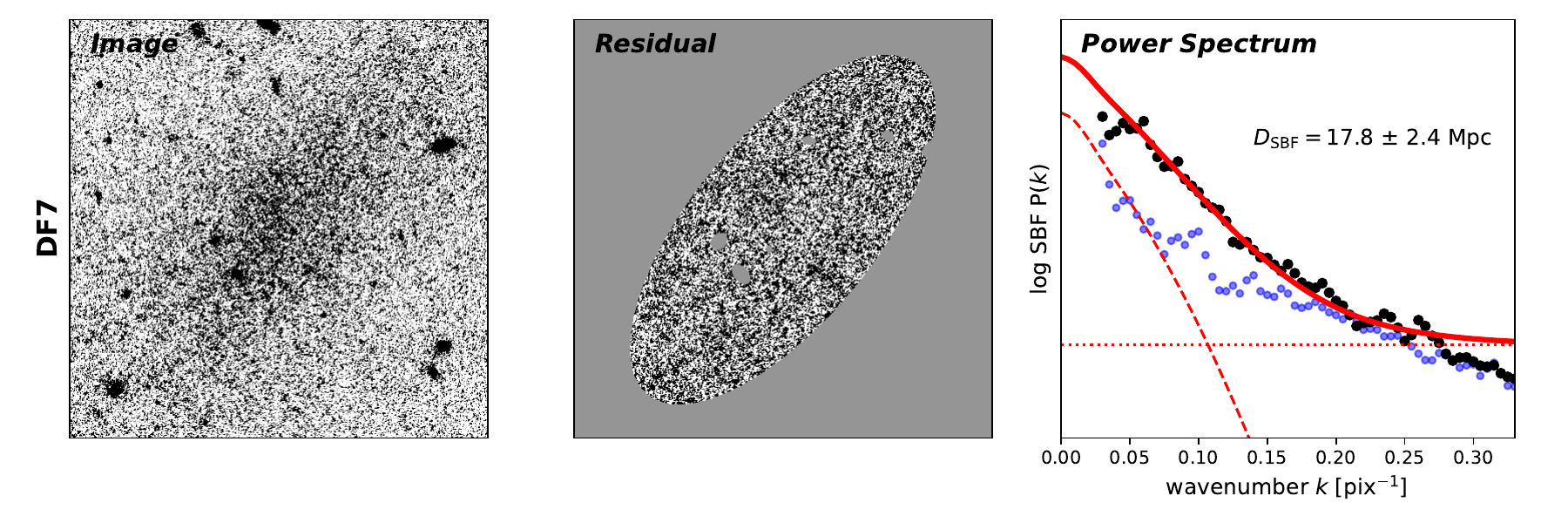}
\caption{Examples of our SBF distance measurements, where DF2 and DF7 (labeled at far left) represent cases of high and low SB, respectively. The left panel shows the original HST/ACS F814W image, and the middle panel is the normalized residual image, with contaminants and the region beyond $1 R_{\rm e}$ masked. The right panel presents the azimuthally averaged power spectrum of the galaxy (black points), overlaid with the best-fitting model (red solid curve) with a combination of the PSF and white-noise components (shown individually with the red dashed and dotted lines, respectively). The contribution from unmasked background sources is shown by the blue points.}
\label{fig:sbf}
\end{figure*}

We perform SBF distance measurements using HST imaging obtained in the programs GO 14644 and 16912 (PI: Pieter van Dokkum). All the trail dwarfs were observed with the Advanced Camera for Surveys (ACS) Wide Field Channel for one orbit each in the F606W and F814W bands. For all programs, we follow a special flat-fielding procedure for ACS, applying corrections to the \texttt{flc} files prior to drizzling, as described in the most recent analyses of DF2 and DF4 by \cite{vanDokkum2022b}. The lanczos3 interpolation kernel is used in drizzling, which is preferred for SBF analysis \citep{Mei2005,Cantiello2005}. Among all 12 trail dwarfs, RCP~17, RCP~21, RCP~28, and RCP~32 are excluded from our analysis in this work because their SBs are too low for reliable SBF measurements. Importantly, the extended halo of NGC~1052 was captured in the ACS pointing of RCP~26, and the outskirts of NGC~1035 fall within the parallel Wide Field Camera 3 (WFC3) field of view of the RCP~17 visit, allowing us to determine the SBF distances of these two massive, potential host galaxies as well. Figure \ref{fig:ngc1052_field} shows a gallery of all the galaxies studied in this work. All the HST data used here can be found in MAST at doi: 10.17909/krax-y793.

Although additional orbits for DF2, DF4, and DF5 were obtained in Programs 15695 and 15851 (PI: Pieter van Dokkum), our analysis here is limited to the single-orbit data, as small misalignments in stacking may smooth the SBF signal. In practice, for our galaxies, the uncertainties in the SBF distances are dominated by color, unresolved background sources, and calibration errors, rather than by the SBF signal of the galaxy light, so stacking would not provide significant improvement. In any case, we have performed SBF distance measurements for every single orbit separately for these three galaxies and found consistent results.

The SBF signal is measured from the F814W images. Compared to F606W, F814W yields a significantly stronger SBF amplitude dominated by bright red giant stars, and provides a more stable empirical calibration for distance measurements. The first step is to model the 2D light distribution of our galaxies, which was done in \cite{Tang2025a} with {\tt GALFIT} \citep{Peng2002} for the dwarf galaxies in our sample. Our modeling combines a single S\'ersic profile and a sky plane, which provides reasonable fits in all cases. To minimize contamination from foreground stars and background galaxies, we run {\tt SExtractor} \citep{Bertin1996} with a low detection threshold of 2$\sigma$ to generate masks. Additionally, the sources with $M_{\rm F814W}<-5$ are masked iteratively during our SBF distance measurements. The morphological parameters are initially fitted in the stacked F606W+F814W images, and applied as fixed values to individual filters thereafter (e.g., to derive galaxy colors). Here, we use the same procedure to model the NGC~1052 halo and the NGC~1035 outskirts, and the fits are satisfactory.

The residual images from {\tt GALFIT} are normalized by dividing by the square root of the smooth S\'ersic models. We mask contaminants with {\tt SExtractor}, and focus only on the region within the one $R_{\rm e}$ ellipse to achieve higher S/N for our SBF measurement. NGC 1052 and NGC 1035 are exceptions, with their regions defined by image boundaries and isophotal ellipses, aiming for higher S/N and less contamination from substructures (see Figure \ref{fig:sbf_all}). We measure the SBF signal by Fourier transforming the residual image and calculating the azimuthally averaged power spectrum. The one-dimensional power spectrum of the point spread function (PSF) is obtained in the same way, with the mask window function convolved. The PSF for each field is built empirically using bright stars. The relation between the residual image power spectrum and the PSF power spectrum is given in \cite{Tonry1988}:
$$P_{\rm image}(k)=(P_{\rm SBF}+P_{\rm BG}) \times P_{\rm PSF}(k)+P_1,$$
where $P_{\rm SBF}$ is the SBF signal from the stars in the galaxy, $P_{\rm BG}$ is the contribution from the unmasked background sources, and $P_1$ is the white-noise term. We fit only over the wavenumber range of $0.05<k<0.30$ pixel$^{-1}$, as the smaller and larger wavenumbers are dominated by residual large-scale structure in the image and by noise correlations, respectively \citep{Mei2005}. We repeat the same analysis on blank fields near each galaxy to estimate the SBF signal contributed by the background, and subtract it from the signals measured on galaxies. The SBF signal uncertainty is derived from the standard deviation of 1000 Monte Carlo realizations, generated by injecting Gaussian noise based on the error map. In these Monte Carlo realizations, the nearby blank background field is randomly selected to account for uncertainties from unresolved background sources. At the same time, we randomly vary the upper wavenumber limit $k$ between 0.25 and 0.30 pixel$^{-1}$ (e.g., \citealt{Carlsten2019,Li2024}). We also correct the SBF signal for Galactic extinction. Examples of our SBF measurement procedure are presented in Figure~\ref{fig:sbf} (the rest of the galaxies are in Appendix~\ref{sbf_all}). 

\setlength{\tabcolsep}{3pt}
\begin{deluxetable*}{lcccccccc}
\tabletypesize{\footnotesize}
\renewcommand{\arraystretch}{1.05}
\tablewidth{0pt}
\tablecaption{Properties of NGC~1052, NGC~1035, and eight dwarf galaxies on the trail studied in the work.
\label{tab:data_properties}}
\tablehead{
\colhead{Galaxy} & \colhead{RA} & \colhead{Dec} & \colhead{Radial Velocity} & \colhead{$g-i$} & \colhead{$\bar{m}_{\rm F814W}$} & \colhead{$D_{\rm SBF, HST}$} & \colhead{$D_{\rm TRGB, HST}$} & \colhead{$D_{\rm TRGB, JWST}$} \\
\colhead{} & \colhead{[deg]} & \colhead{[deg]} & \colhead{[km s${}^{-1}$]} & \colhead{[mag]} & \colhead{[mag]} & \colhead{[Mpc]} & \colhead{[Mpc]} & \colhead{[Mpc]}}
\startdata
    NGC 1052 & 40.26999 & $-$8.25576 & $1488_{-5}^{+5}$ & $1.07 \pm 0.01$ & $30.23 \pm 0.03$ & $21.3 \pm 1.7$ & -- & -- \\
    NGC 1035 & 39.87145 & $-$8.13328 & $1288_{-4}^{+4}$ & $0.85 \pm 0.01$ & $29.04 \pm 0.03$ & $13.8 \pm 1.1$ & -- & -- \\
    \hline
    DF2 (RCP 29)& 40.44531 & $-$8.40297 & $1805_{-1}^{+1}$ & $0.86 \pm 0.01$ & $29.65 \pm 0.03$ & $17.7 \pm 1.4$ & $21.7\pm 1.2$ & $17.6 \pm 0.6$ \\
    RCP 26 & 40.28970 & $-$8.29691 & $1670_{-10}^{+10}$ & $0.93 \pm 0.03$ & $29.87 \pm 0.20$ & $18.9 \pm 2.2$ & -- & -- \\
    DF9 & 40.02927 & $-$8.22902 & $1648_{-1}^{+1}$ & $0.81 \pm 0.01$ & $29.71 \pm 0.05$ & $18.7 \pm 1.5$ & -- & -- \\
    RCP 16 (Ta21-12000) & 39.91385 & $-$8.22845 & $1540_{-8}^{+9}$ & $0.77 \pm 0.05$ & $29.99 \pm 0.19$ & $21.8 \pm 2.6$ & -- & -- \\
    DF4 (RCP 12) & 39.81271 & $-$8.11597 & $1433.3_{-0.4}^{+0.3}$ & $0.87 \pm 0.01$ & $29.79 \pm 0.06$ & $18.8 \pm 1.7$ & $20.0 \pm 1.6$ & -- \\
    DF5 (RCP 11) & 39.80284 & $-$8.14083 & $1440_{-20}^{+30}$ & $0.67 \pm 0.04$ & $29.74 \pm 0.22$ & $20.5 \pm 2.7$ & -- & -- \\
    LEDA 4014647 & 39.70206 & $-$8.04938 & $1596_{-2}^{+3}$ & $0.81 \pm 0.01$ & $29.58 \pm 0.07$ & $17.7 \pm 1.4$ & -- & -- \\
    DF7 (RCP 9) & 39.62401 & $-$7.92576 & $1690_{-10}^{+10}$ & $0.77 \pm 0.11$ & $29.55 \pm 0.21$ & $17.8 \pm 2.4$ & -- & -- \\
\enddata
\tablecomments{Radial velocities for NGC~1052 and NGC~1035 are taken from NASA/IPAC Extragalactic Database (NED; from \citealt{Fouque1992} and \citealt{vanDriel2016}, respectively), and those for the trail dwarfs are from \cite{Keim2025}. The SBF magnitudes $\bar{m}_{\rm F814W}$ reported here have been corrected for Galactic extinction and for contributions from unmasked background sources. The $g-i$ colors are reported in the Subaru/HSC system, except for NGC~1052, for which the F475W$-$F814W color is used instead. The TRGB distances for DF2 and DF4 based on HST are sourced from \cite{Shen2021b,Shen2023} and \cite{Danieli2020}, respectively.}
\end{deluxetable*}

Converting the SBF signals in the F814W filter into distances for the trail dwarfs requires a reliable calibration for blue colors, which was not available in the earliest SBF studies of the NGC~1052 group \citep{vanDokkum2018a,Cohen2018,Trujillo2019,Blakeslee2018}.
Now we adopt the calibration derived by \cite{Kim2021}, which was built in the magnitude system of Hyper Suprime-Cam (HSC) on the Subaru Telescope. Using the best-fit spectral energy distributions of the trail dwarfs in \cite{Tang2025a}, we obtain the best estimated $(g-i)_{\rm HSC}$ colors and the transformation between F814W and $i_{\rm HSC}$ filters for each galaxy.

For NGC~1052, we instead apply the calibration for red massive galaxies from \cite{Blakeslee2010}. Using {\tt FSPS} \citep{Conroy2009}, its $\rm (F606W-F814W)_0$ color (0.55 mag) can be transformed to $\rm (F475W-F814W)_0$ color (1.07 mag) as required in the calibration.
The case of NGC~1035 requires a hybrid approach. As a giant galaxy with a bluer color, it falls outside the standard color range of the \cite{Blakeslee2010} calibration. Fortunately, we find that the \cite{Kim2021} and \cite{Blakeslee2010} relations yield consistent results when extrapolated into each other's color regimes. Therefore, we estimate the distance for NGC~1035 using the \cite{Kim2021} calibration by transforming its $\rm (F606W-F814W)_0$ color (0.44 mag) into a $(g-i)_{\rm HSC, 0}$ color (0.85 mag). We note that the colors of NGC~1052 and NGC~1035 correspond to the same regions where the SBF signals are measured (see Figure \ref{fig:sbf_all}).

We note that using the extrapolation of the \cite{Blakeslee2010} calibration for NGC~1035 would move this galaxy about 1 Mpc farther. Also, we find consistent results for all galaxies if we adopt the theoretical calibration by \cite{Greco2021}, while using the empirical calibration for dwarf galaxies from \cite{Carlsten2019} would systematically shift the redder half of our dwarfs $\sim 1$ Mpc closer. However, these uncertainties would not substantially affect our results and conclusions.

The measured SBF distances are listed in Table \ref{tab:data_properties}. This total uncertainty incorporates contributions from the measurement noise of the SBF signal from the galaxies and backgrounds, the errors of the photometric colors, and the systematic uncertainty of the calibrations. The discrepancy among different empirical calibrations is not included in the reported uncertainties.

\section{TRGB distance of DF2 From JWST} \label{sec:trgb}

\begin{figure*}
\centering
\includegraphics[width=0.95\textwidth]{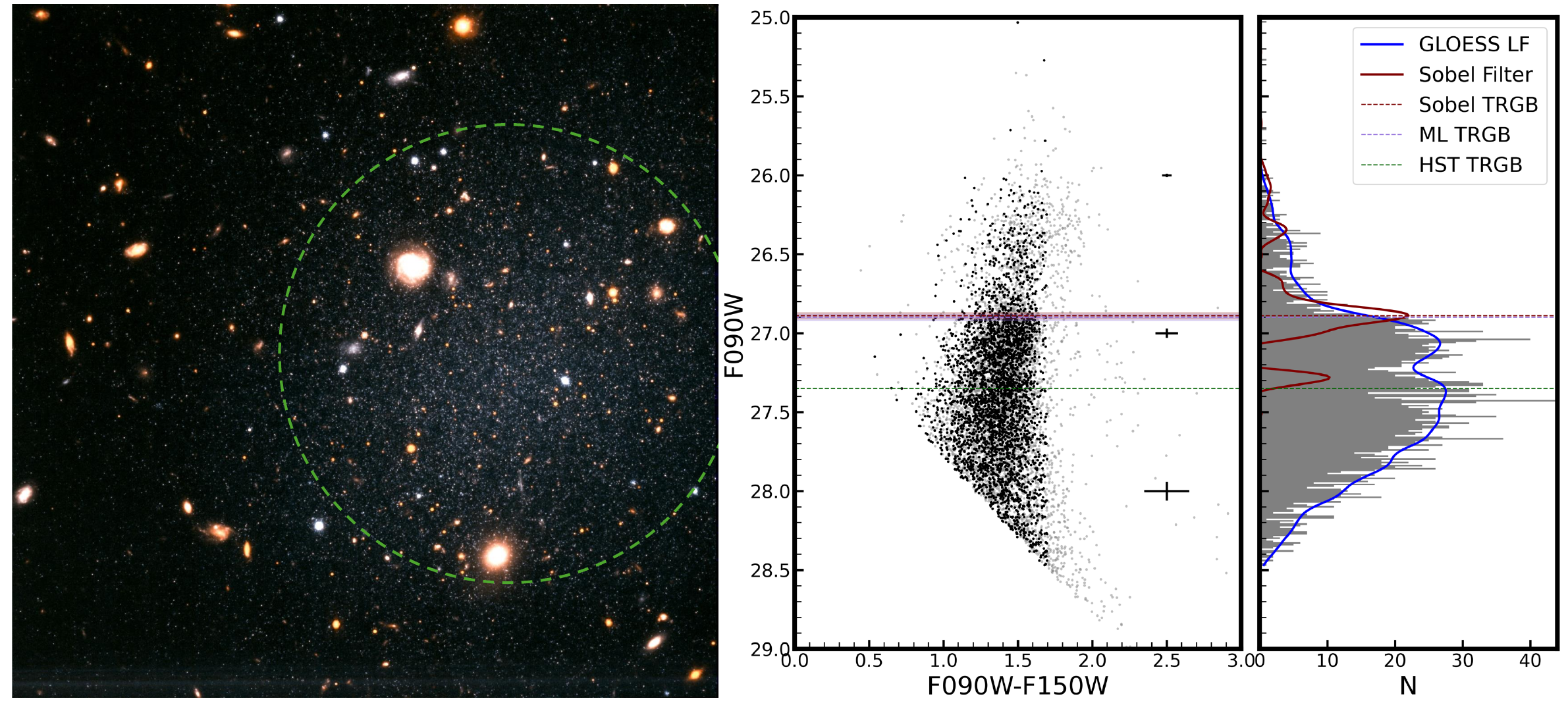}
\caption{\textbf{Left:} Pseudo-color (F090W/F090W+F444W/F444W) JWST NIRCam image of NGC~1052-DF2. The dashed green circle denotes $R_\mathrm{e}$. \textbf{Middle:} CMD built from the available F090W+F150W data. Black points denote stars retained for the edge-detection analysis, whereas the fainter gray points were removed with spatial and/or color cuts. Representative photometric uncertainties, calculated at the median color of the TRGB (F090W$-$F150W = 1.4~mag), are shown as black crosses. \textbf{Right:} Luminosity function of stars used for the TRGB analysis (gray). The GLOESS-smoothed version is shown in blue, and results of the Sobel edge-detection filter are given in red. TRGB magnitude results from a maximum-likelihood technique are shown as a purple horizontal dashed line, and agree within uncertainties with the results from the edge-detection (red dashed line). The implied magnitude of the F090W TRGB from the HST TRGB distance \citep{Shen2023} is shown in green, and is in clear disagreement with the JWST data. The JWST photometry catalog is available on Zenodo doi: \href{https://doi.org/10.5281/zenodo.20740394}{10.5281/zenodo.20740394}.}
\label{fig:df2_trgb}
\end{figure*}

DF2 was observed serendipitously via a JWST pure parallel survey \citep{Morishita2023}. These observations were tied to primary observations of a runaway supermassive black hole \citep{vanDokkum2026}, which itself was a chance discovery resulting from prior HST observations of the NGC~1052 system \citep{vanDokkum2023}. The JWST/NIRCam observations include data in F090W and F444W (12626~s each), as well as F150W and F356W (1031~s each). Notably, the two short-wavelength filters are particularly useful for building high-precision color--magnitude diagrams (CMDs) of stellar populations and measuring the brightness of the TRGB \citep{Weisz2023, Anand2025}. While the F444W imaging is much deeper than the F150W data, it suffers from significantly worse resolution and has not been recommended for TRGB work \citep{Newman2024_jwst}. We performed PSF photometry on this dataset with {\tt DOLPHOT}'s JWST module, adopting the recommended parameters for NIRCam data \citep{Dolphin2016, Weisz2024} and using the Vega photometric zeropoints (as opposed to the Sirius--Vega variant). The initial photometric catalog was trimmed using the following quality cuts to retain only high-fidelity sources: \texttt{crowd} $<$ 0.5, $\texttt{sharpness}^{2}$ $\le$ 0.01, \texttt{type} $\le$ 2, \texttt{S/N} $\ge$ 5, and \texttt{flag} $\le$ 2 \citep{Warfield2023, Anand2025}. All of these selections were applied to the photometry from both F090W and F150W.
A pseudo-color image of the field and the resulting color--magnitude diagram after the specified quality cuts to the data are shown in Figure \ref{fig:df2_trgb}. 

Notably, the F090W TRGB is insensitive to metallicity and age effects, with no statistically significant trend with color out to a red color limit of F090W$-$F150W = 1.68 mag \citep{Newman2024_jwst}. This allows us to simply measure the TRGB feature without an explicit color correction of the CMD, as is required at longer wavelengths \citep{Newman2024_hst}. We use a Sobel edge-detection algorithm to isolate the TRGB feature \citep{Lee1993}. We first bin the underlying luminosity (LF) function blueward of (F090W$-$F150W)$_{0}$ = 1.68 mag with a bin width of 0.01~mag, and smooth this LF with a Gaussian-windowed, Locally Weighted Scatterplot Smoothing (GLOESS) algorithm \citep{Beaton2019} to reduce the effects of noise. Then, we use a Sobel filter on the smoothed LF to find a value of $m_{\rm TRGB}$ = 26.88 $\pm$ 0.02~mag, where the uncertainty is based on the simulations of \cite{Madore2023}. As a first test for the effects of potential photometric bias resulting from stellar crowding, we repeat this same measurement after removing stars within the inner 20$''$ ($\approx R_{\rm e}$) of the imaging (see the left-hand side of Figure \ref{fig:df2_trgb}) to find $m_{\rm TRGB}$ = 26.89 $\pm$ 0.02~mag, indicative of minimal-to-no apparent photometric bias at the level of the TRGB.

As a further test of the edge-detection routine, we also perform measurements with a maximum-likelihood algorithm which explicitly takes into account the observed photometric bias, completeness, and errors \citep{Makarov2006}. We perform artificial star experiments with {\tt DOLPHOT} (see Appendix \ref{art_stars}) and use these within the maximum-likelihood routine to determine $m_{\rm TRGB}$ = 26.90 $\pm$ 0.02~mag, fully consistent with the edge-detection method. Taking into account the known foreground extinction \citep{Schlafly2011} of $E(B-V)$ = 0.021~mag (corresponding to $A_{\rm F090W}$ = 0.03~mag), we find $m_{\rm TRGB,0}$ = 26.87 $\pm$ 0.02~mag. Besides the basic measurement error, we follow \cite{Anand2024} and add in quadrature the uncertainties in {\tt DOLPHOT}'s aperture correction routine (0.02~mag), the foreground reddening (0.01~mag), the PSF stability of NIRCam (0.01~mag), and any minor residual color effects (0.01~mag), yielding an overall uncertainty of $\pm$0.033~mag. We also note that the median color of stars (F090W$-$F150W = 1.43~mag) within 0.05~mag fainter than the measured F090W TRGB is consistent with expectations for old, metal-poor stars \citep{Anand2024, Newman2024_jwst}, further supporting this new JWST TRGB measurement.

To arrive at a distance and final uncertainty, we use the absolute TRGB calibration of $M_{\rm TRGB}$ = $-$4.36 $\pm$ 0.056~mag derived from JWST observations in the same passbands of the megamaser host NGC~4258 \citep{Anand2024}. This F090W calibration agrees with the two others in the literature \citep{Newman2024_jwst,Freedman2025} to within $\sim$1$\%$ in distance. Adding the prior statistical uncertainty in quadrature with the zero-point uncertainty, we arrive at a distance modulus of $\mu$ = 31.23 $\pm$ 0.07~mag, or distance of $D$ = 17.6 $\pm$ 0.6~Mpc.

\section{Results} \label{sec:results}

\begin{figure*}
\centering
\includegraphics[width=1\textwidth]{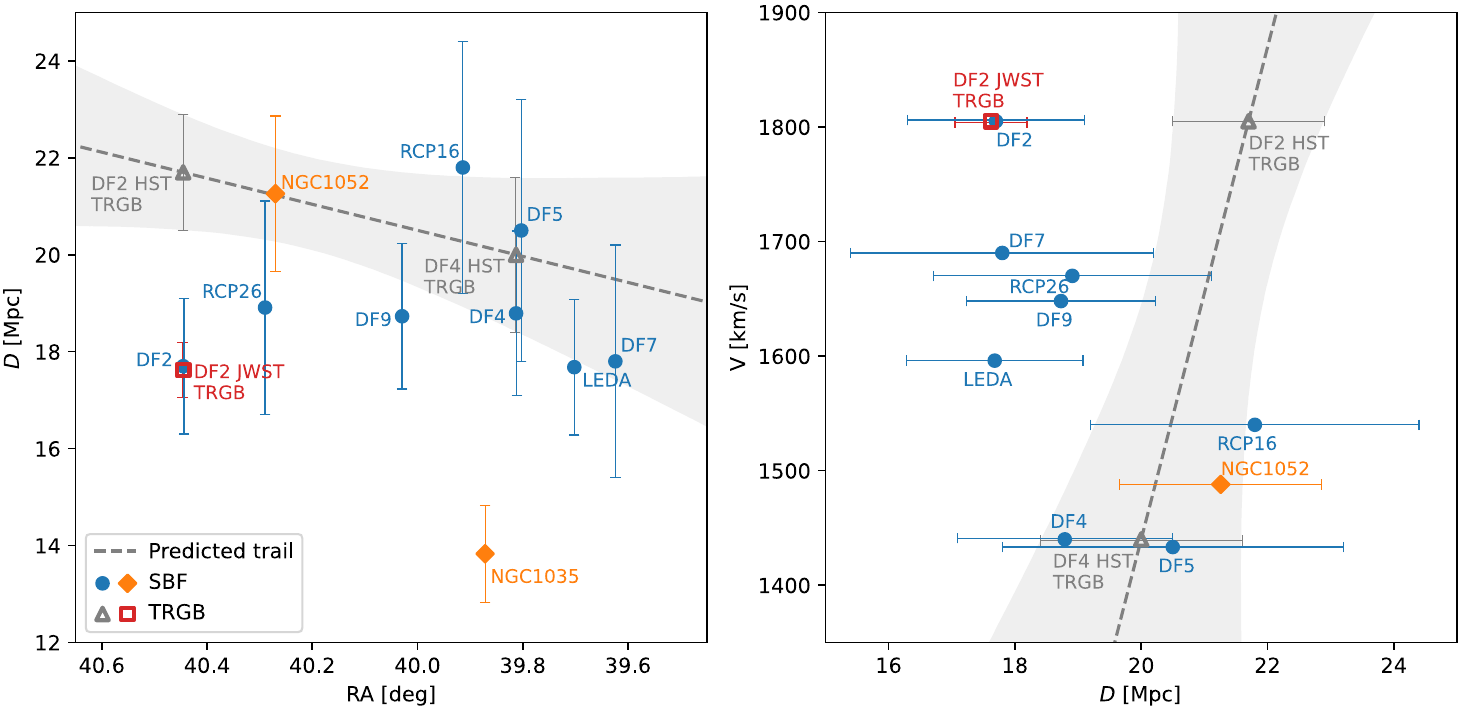}
\caption{Correlation between galaxy distance and Right Ascension (Left), and between distance and radial velocity (Right). The blue solid circles for the eight trail dwarfs and the orange solid diamonds for NGC~1052 and NGC~1035 show their SBF distances derived from HST imaging. The gray open triangles mark the HST-based TRGB distances of DF2 and DF4, which define a predicted trail (gray dashed line and shaded band). The TRGB distance of DF2 obtained from JWST is marked as a red open square.}
\label{fig:results}
\end{figure*}

We show the SBF distance distribution of the trail dwarfs, NGC~1052, and NGC~1035 in Figure \ref{fig:results}. Our distance estimates for NGC~1052 and NGC~1035 are consistent with literature values listed in NED based on various methods, although \cite{Jacoby2024} reported a closer distance for NGC~1052 of $17.9_{-0.6}^{+0.3}$ Mpc based on the planetary nebula LF. For DF2, our SBF distance aligns with previous results derived from the same HST observation \citep{Blakeslee2018,Cohen2018,vanDokkum2018d,Zonoozi2021}, while being significantly closer than the TRGB distance obtained from deep HST imaging \citep{Shen2021b,Shen2023}. In the case of DF4, our SBF distance is in good agreement with the HST-based TRGB distance \citep{Danieli2020} and with the previous SBF measurement \citep{Cohen2018}. \cite{Cohen2018} reported much closer distances for DF5 and DF7, but did not account for the background contribution to the SBF signal, which is critical for galaxies with extremely low SB.

Importantly, since we have performed the SBF distances of these galaxies in a homogeneous way, we can be particularly confident in the relative distances between them. Within the uncertainty, none of the trail dwarfs agree with the distance of NGC~1035 at $\sim$14 Mpc. Therefore it is highly unlikely for the trail to be an artifact produced by the projection of overlapping satellites from the NGC~1052 and NGC~1035 galaxy groups.

All of the trail dwarfs cluster around 20 Mpc, consistent with the distance of NGC~1052, but there is a tendency for more dwarfs to lie at closer distances.
Their distances show no significant correlation with the Right Ascension (which is roughly parallel to the trail). However, given the uncertainties, we cannot rule out the possibility that all these dwarfs lie almost exactly on a single straight line in this coordinate, with a slope consistent with that implied by the HST TRGB distances of DF2 and DF4 in the bullet dwarf scenario. Therefore, based on the SBF distances alone, we cannot determine whether these dwarfs constitute a remarkable linear structure or are simply random members of the NGC 1052 group.

Figure \ref{fig:results} also shows the correlation between distance and radial velocity for our galaxies. The bullet dwarf collision scenario predicts that galaxies along the trail should follow a linear relation in this phase space \citep{Lee2024}, whereas normal group members would exhibit a more random velocity distribution. Within the uncertainties of our SBF distances, however, we cannot distinguish between these two scenarios. Nevertheless, the strong correlation between velocity and Right Ascension found by \cite{Keim2025} already provides high statistical significance in support of a physically coherent trail.

The TRGB distance of $17.6 \pm 0.6$ Mpc derived from JWST reaffirms that DF2 is not located at a much closer distance of $\sim 13$ Mpc, agreeing with our SBF distance of $17.7 \pm 1.4$ Mpc and the $18.7 \pm 1.7$ Mpc distance derived from an empirical megamaser\,--\,TRGB\,--\,SBF distance ladder \citep{vanDokkum2018d}.\footnote{Our TRGB result is also consistent with the 16.2 $\pm$ 1.3 Mpc distance inferred \cite{Beasley2025} from GC internal velocity dispersions, assuming the DF2 GCs follow the same scaling relations as Milky Way and M31 GCs (e.g., allowing for larger sizes of the DF2 GCs was estimated to shift the distance to $\sim$17-18 Mpc).}
It is, however, significantly closer than the $21.7 \pm 1.2$ Mpc distance obtained from a TRGB measurement based on very deep HST imaging\footnote{We note that the HST CMD of DF2 in \cite{Shen2021b} shows a modest population of sources blueward of the RGB. These sources are likely background galaxies, as shown by \cite{Jang2017} and \cite{Anand2025b}. With the improved spatial resolution of JWST data, they can be more readily identified and removed. In addition, they may exhibit relatively red F090W$-$F150W colors that overlap with the RGB stars.}
\citep{Shen2023}.
We have explored possible origins for this discrepancy, such as errors in photometric zeropoints, image alignments, and methodological differences, with no satisfactory conclusions. In particular, we find $\langle {\rm F814W} - {\rm F090W}\rangle = 0.22 \pm 0.03$ mag for DF2's GCs, consistent with the expected value for old, metal-poor stellar populations ($\approx 0.19$ mag). However, for stars near the tip of the RGB we find ${\rm F814W} - {\rm F090W} \approx 0.5$ mag, much redder than the expected value of $\approx 0.3$ mag. Given the much higher S/N ratio of stars in the JWST data, the consistency between JWST and SBF measurements, and the fact that $\sim 20$ Mpc has been shown to be the practical limit for HST TRGB measurements \citep{Anand2022}, we cautiously infer that the true distance of DF2 is $\sim 18$ Mpc rather than $\sim 22$ Mpc.

\section{Summary \& Discussion} \label{sec:discussion}

In this Letter, we present new SBF distance measurements for eight dwarf galaxies along the remarkable linear trail in the NGC~1052 field, as well as for the two giant galaxies NGC~1052 and NGC~1035. We find that all the trail dwarfs are located at $\sim$20 Mpc, consistent with the distance of NGC~1052, indicating they are not associated with the foreground NGC~1035 group at $\sim$13 Mpc. This result suggests that the trail is unlikely to be an artifact arising from the projection of the overlapping satellites of the NGC~1052 and NGC~1035 groups. However, given the relatively large uncertainties inherent to the SBF method, we cannot yet definitively confirm the existence of the linear structure in three-dimensional space.

We obtain a TRGB distance of 17.6 $\pm$ 0.6~Mpc for DF2 based on JWST observations, which is still substantially more distant than that of the NGC~1035 group and continues to support the conclusion that DF2 is deficient in DM. This value is consistent with our SBF distance, but significantly closer than the TRGB distance of $21.7 \pm 1.2$ Mpc derived from deep HST imaging -- a discrepancy that remains puzzling. We highlight the importance of obtaining JWST imaging for other trail dwarfs to measure their TRGB distances with high S/N without introducing problems from measuring resolved star photometry on individual shallow exposures. Fortunately, our new JWST program has been approved to pursue this goal (GO 12429, PI: A.~J.~Romanowsky). Confirming the distances to DF4 and DF9 is particularly important, as their relative line-of-sight separation from DF2 could have major implications for how we interpret their formation. Given that DF2 has a larger radial velocity than DF4 and DF9, if DF2 is indeed closer, tracing their motions backward would imply that their separations were larger in the past rather than smaller. This may pose a challenge to the bullet dwarf scenario, unless the orbit of one or more of these galaxies was dramatically modified through gravitational interactions with a nearby massive object after the collision. In this context, it is noteworthy that DF4 does exhibit tidal features, whereas DF2 remains under debate \citep{Keim2022,Golini2024}. Numerical simulations can further test whether or not viable initial conditions for a bullet dwarf collision exist that are consistent with the DF2 distance inferred from the JWST data, similar to the approach of \cite{Lee2024}.

Another piece to the puzzle is the systematic offset shown between DF2 and DF4 in the HST TRGB measurements, such that DF4 is $1.7\pm0.5$ Mpc closer than DF2. This measurement may be more secure than the absolute distance. In that case, DF4 may be at $\sim$ 16.5 Mpc and the trail may be on the near side of the NGC~1052 group rather than the far side. We note that the constraints on the dark matter content of the trail galaxies are not significantly impacted by the range of distances discussed here (see Fig.\ 6 in \citealt{Keim2026}).
Also, the GCs in DF2 are still overluminous with the closer distances we obtained, similar to the 19 Mpc distance originally used in the GCs study by \citet{vanDokkum2018b}.

We emphasize that there is currently no alternative theoretical model that can simultaneously explain the distinctive properties of the trail dwarfs -- most notably the three confirmed DM-deficient galaxies in the same field, the striking similarities between DF2 and DF4, the kinematic connection of the trail members, and their alignment in position angle. Follow-up measurements of high-precision distance and DM content in the future will provide definitive tests of the bullet dwarf collision scenario and of the existence of the trail structure.

\begin{acknowledgments}

We thank the referee for useful feedback. AJR was supported by National Science Foundation grant AST-2308390. Support from STScI grant HST-GO-16912 is gratefully acknowledged.

\end{acknowledgments}

\bibliography{reference}{}

@ARTICLE{Anand2022,
       author = {{Anand}, Gagandeep S. and {Tully}, R. Brent and {Rizzi}, Luca and {Riess}, Adam G. and {Yuan}, Wenlong},
        title = "{Comparing Tip of the Red Giant Branch Distance Scales: An Independent Reduction of the Carnegie-Chicago Hubble Program and the Value of the Hubble Constant}",
      journal = {\apj},
         year = 2022,
        month = jun,
       volume = {932},
       number = {1},
          eid = {15},
        pages = {15},
          doi = {10.3847/1538-4357/ac68df},
archivePrefix = {arXiv},
       eprint = {2108.00007},
 primaryClass = {astro-ph.CO},
       adsurl = {https://ui.adsabs.harvard.edu/abs/2022ApJ...932...15A}
}

@ARTICLE{Anand2024,
       author = {{Anand}, Gagandeep S. and {Riess}, Adam G. and {Yuan}, Wenlong and {Beaton}, Rachael and {Casertano}, Stefano and {Li}, Siyang and {Makarov}, Dmitry I. and {Makarova}, Lidia N. and {Tully}, R. Brent and {Anderson}, Richard I. and {Breuval}, Louise and {Dolphin}, Andrew and {Karachentsev}, Igor D. and {Macri}, Lucas M. and {Scolnic}, Daniel},
        title = "{Tip of the Red Giant Branch Distances with JWST: An Absolute Calibration in NGC 4258 and First Applications to Type Ia Supernova Hosts}",
      journal = {\apj},
         year = 2024,
        month = may,
       volume = {966},
       number = {1},
          eid = {89},
        pages = {89},
          doi = {10.3847/1538-4357/ad2e0a},
archivePrefix = {arXiv},
       eprint = {2401.04776},
 primaryClass = {astro-ph.CO},
       adsurl = {https://ui.adsabs.harvard.edu/abs/2024ApJ...966...89A}
}

@ARTICLE{Anand2025,
       author = {{Anand}, Gagandeep S. and {Tully}, R. Brent and {Cohen}, Yotam and {Shaya}, Edward J. and {Makarov}, Dmitry I. and {Makarova}, Lidia N. and {Chazov}, Maksim I. and {Blakeslee}, John P. and {Cantiello}, Michele and {Jensen}, Joseph B. and {Kourkchi}, Ehsan and {Raimondo}, Gabriella},
        title = "{The TRGB{\textendash}SBF Project. II. Resolving the Virgo Cluster with JWST}",
      journal = {\apj},
         year = 2025,
        month = mar,
       volume = {982},
       number = {1},
          eid = {26},
        pages = {26},
          doi = {10.3847/1538-4357/adb399},
archivePrefix = {arXiv},
       eprint = {2408.16810},
 primaryClass = {astro-ph.GA},
       adsurl = {https://ui.adsabs.harvard.edu/abs/2025ApJ...982...26A}
}

@ARTICLE{Anand2025b,
       author = {{Anand}, Gagandeep S. and {Ben{\'\i}tez-Llambay}, Alejandro and {Beaton}, Rachael and {Fox}, Andrew J. and {Navarro}, Julio F. and {D'Onghia}, Elena},
        title = "{The First RELHIC? Cloud-9 is a Starless Gas Cloud}",
      journal = {\apjl},
         year = 2025,
        month = nov,
       volume = {993},
       number = {2},
          eid = {L55},
        pages = {L55},
          doi = {10.3847/2041-8213/ae1584},
archivePrefix = {arXiv},
       eprint = {2508.20157},
 primaryClass = {astro-ph.GA},
       adsurl = {https://ui.adsabs.harvard.edu/abs/2025ApJ...993L..55A}
}

@ARTICLE{Beasley2025,
       author = {{Beasley}, M.~A. and {Fahrion}, K. and {Guerra Arencibia}, S. and {Gvozdenko}, A. and {Montes}, M.},
        title = "{A new way to measure the distance to NGC1052-DF2}",
      journal = {\aap},
         year = 2025,
        month = may,
       volume = {697},
          eid = {A144},
        pages = {A144},
          doi = {10.1051/0004-6361/202452446},
archivePrefix = {arXiv},
       eprint = {2503.03403},
 primaryClass = {astro-ph.GA},
       adsurl = {https://ui.adsabs.harvard.edu/abs/2025A&A...697A.144B}
}

@ARTICLE{Beaton2019,
       author = {{Beaton}, Rachael L. and {Seibert}, Mark and {Hatt}, Dylan and {Freedman}, Wendy L. and {Hoyt}, Taylor J. and {Jang}, In Sung and {Lee}, Myung Gyoon and {Madore}, Barry F. and {Monson}, Andrew J. and {Neeley}, Jillian R. and {Rich}, Jeffrey A. and {Scowcroft}, Victoria},
        title = "{The Carnegie-Chicago Hubble Program. VII. The Distance to M101 via the Optical Tip of the Red Giant Branch Method}",
      journal = {\apj},
         year = 2019,
        month = nov,
       volume = {885},
       number = {2},
          eid = {141},
        pages = {141},
          doi = {10.3847/1538-4357/ab4263},
archivePrefix = {arXiv},
       eprint = {1908.06120},
 primaryClass = {astro-ph.GA},
       adsurl = {https://ui.adsabs.harvard.edu/abs/2019ApJ...885..141B}
}

@ARTICLE{Bertin1996,
       author = {{Bertin}, E. and {Arnouts}, S.},
        title = "{SExtractor: Software for source extraction.}",
      journal = {\aaps},
         year = 1996,
        month = jun,
       volume = {117},
        pages = {393-404},
          doi = {10.1051/aas:1996164},
       adsurl = {https://ui.adsabs.harvard.edu/abs/1996A&AS..117..393B}
}

@ARTICLE{Blakeslee2010,
       author = {{Blakeslee}, John P. and {Cantiello}, Michele and {Mei}, Simona and {C{\^o}t{\'e}}, Patrick and {Barber DeGraaff}, Regina and {Ferrarese}, Laura and {Jord{\'a}n}, Andr{\'e}s and {Peng}, Eric W. and {Tonry}, John L. and {Worthey}, Guy},
        title = "{Surface Brightness Fluctuations in the Hubble Space Telescope ACS/WFC F814W Bandpass and an Update on Galaxy Distances}",
      journal = {\apj},
         year = 2010,
        month = nov,
       volume = {724},
       number = {1},
        pages = {657-668},
          doi = {10.1088/0004-637X/724/1/657},
archivePrefix = {arXiv},
       eprint = {1009.3270},
 primaryClass = {astro-ph.CO},
       adsurl = {https://ui.adsabs.harvard.edu/abs/2010ApJ...724..657B}
}

@ARTICLE{Blakeslee2018,
       author = {{Blakeslee}, John P. and {Cantiello}, Michele},
        title = "{Independent Analysis of the Distance to NGC{\,}1052-DF2}",
      journal = {Research Notes of the American Astronomical Society},
         year = 2018,
        month = aug,
       volume = {2},
       number = {3},
          eid = {146},
        pages = {146},
          doi = {10.3847/2515-5172/aad90e},
archivePrefix = {arXiv},
       eprint = {1808.02176},
 primaryClass = {astro-ph.GA},
       adsurl = {https://ui.adsabs.harvard.edu/abs/2018RNAAS...2..146B}
}

@ARTICLE{Buzzo2022,
       author = {{Buzzo}, Maria Luisa and {Forbes}, Duncan A. and {Brodie}, Jean P. and {Romanowsky}, Aaron J. and {Cluver}, Michelle E. and {Jarrett}, Thomas H. and {Laine}, Seppo and {Couch}, Warrick J. and {Gannon}, Jonah S. and {Ferr{\'e}-Mateu}, Anna and {Okabe}, Nobuhiro},
        title = "{The stellar populations of quiescent ultra-diffuse galaxies from optical to mid-infrared spectral energy distribution fitting}",
      journal = {\mnras},
         year = 2022,
        month = dec,
       volume = {517},
       number = {2},
        pages = {2231-2250},
          doi = {10.1093/mnras/stac2442},
archivePrefix = {arXiv},
       eprint = {2208.11819},
 primaryClass = {astro-ph.GA},
       adsurl = {https://ui.adsabs.harvard.edu/abs/2022MNRAS.517.2231B}
}

@ARTICLE{Buzzo2025,
       author = {{Buzzo}, Maria Luisa and {Forbes}, Duncan A. and {Romanowsky}, Aaron J. and {Haacke}, Lydia and {Gannon}, Jonah S. and {Tang}, Yimeng and {Hilker}, Michael and {Ferr{\'e}-Mateu}, Anna and {Janssens}, Steven R. and {Brodie}, Jean P. and {Valenzuela}, Lucas M.},
        title = "{A new class of dark matter-free dwarf galaxies?: I. Clues from FCC 224, NGC 1052-DF2, and NGC 1052-DF4}",
      journal = {\aap},
         year = 2025,
        month = mar,
       volume = {695},
          eid = {A124},
        pages = {A124},
          doi = {10.1051/0004-6361/202453522},
archivePrefix = {arXiv},
       eprint = {2502.05405},
 primaryClass = {astro-ph.GA},
       adsurl = {https://ui.adsabs.harvard.edu/abs/2025A&A...695A.124B}
}

@ARTICLE{Buzzo2026,
       author = {{Buzzo}, Maria Luisa and {van Dokkum}, Pieter and {Hilker}, Michael and {Forbes}, Duncan A. and {Romanowsky}, Aaron J. and {Tang}, Yimeng},
        title = "{Dark matter-deficient twins: FCC 224 and FCC 240 as possible analogues of NGC 1052-DF2 and DF4}",
      journal = {arXiv e-prints},
         year = 2026,
        month = may,
          eid = {arXiv:2605.24099},
        pages = {arXiv:2605.24099},
          doi = {10.48550/arXiv.2605.24099},
archivePrefix = {arXiv},
       eprint = {2605.24099},
 primaryClass = {astro-ph.GA},
       adsurl = {https://ui.adsabs.harvard.edu/abs/2026arXiv260524099B}
}

@ARTICLE{Cantiello2005,
       author = {{Cantiello}, Michele and {Blakeslee}, John P. and {Raimondo}, Gabriella and {Mei}, Simona and {Brocato}, Enzo and {Capaccioli}, Massimo},
        title = "{Detection of Radial Surface Brightness Fluctuations and Color Gradients in Elliptical Galaxies with the Advanced Camera for Surveys}",
      journal = {\apj},
         year = 2005,
        month = nov,
       volume = {634},
       number = {1},
        pages = {239-257},
          doi = {10.1086/491694},
archivePrefix = {arXiv},
       eprint = {astro-ph/0507699},
 primaryClass = {astro-ph},
       adsurl = {https://ui.adsabs.harvard.edu/abs/2005ApJ...634..239C}
}

@ARTICLE{Carlsten2019,
       author = {{Carlsten}, Scott G. and {Beaton}, Rachael L. and {Greco}, Johnny P. and {Greene}, Jenny E.},
        title = "{Using Surface Brightness Fluctuations to Study Nearby Satellite Galaxy Systems: Calibration and Methodology}",
      journal = {\apj},
         year = 2019,
        month = jul,
       volume = {879},
       number = {1},
          eid = {13},
        pages = {13},
          doi = {10.3847/1538-4357/ab22c1},
archivePrefix = {arXiv},
       eprint = {1901.07575},
 primaryClass = {astro-ph.GA},
       adsurl = {https://ui.adsabs.harvard.edu/abs/2019ApJ...879...13C}
}

@ARTICLE{Cohen2018,
       author = {{Cohen}, Yotam and {van Dokkum}, Pieter and {Danieli}, Shany and {Romanowsky}, Aaron J. and {Abraham}, Roberto and {Merritt}, Allison and {Zhang}, Jielai and {Mowla}, Lamiya and {Kruijssen}, J.~M. Diederik and {Conroy}, Charlie and {Wasserman}, Asher},
        title = "{The Dragonfly Nearby Galaxies Survey. V. HST/ACS Observations of 23 Low Surface Brightness Objects in the Fields of NGC 1052, NGC 1084, M96, and NGC 4258}",
      journal = {\apj},
         year = 2018,
        month = dec,
       volume = {868},
       number = {2},
          eid = {96},
        pages = {96},
          doi = {10.3847/1538-4357/aae7c8},
archivePrefix = {arXiv},
       eprint = {1807.06016},
 primaryClass = {astro-ph.GA},
       adsurl = {https://ui.adsabs.harvard.edu/abs/2018ApJ...868...96C}
}

@ARTICLE{Conroy2009,
       author = {{Conroy}, Charlie and {Gunn}, James E. and {White}, Martin},
        title = "{The Propagation of Uncertainties in Stellar Population Synthesis Modeling. I. The Relevance of Uncertain Aspects of Stellar Evolution and the Initial Mass Function to the Derived Physical Properties of Galaxies}",
      journal = {\apj},
         year = 2009,
        month = jul,
       volume = {699},
       number = {1},
        pages = {486-506},
          doi = {10.1088/0004-637X/699/1/486},
archivePrefix = {arXiv},
       eprint = {0809.4261},
 primaryClass = {astro-ph},
       adsurl = {https://ui.adsabs.harvard.edu/abs/2009ApJ...699..486C}
}

@ARTICLE{Danieli2019,
       author = {{Danieli}, Shany and {van Dokkum}, Pieter and {Conroy}, Charlie and {Abraham}, Roberto and {Romanowsky}, Aaron J.},
        title = "{Still Missing Dark Matter: KCWI High-resolution Stellar Kinematics of NGC1052-DF2}",
      journal = {\apjl},
         year = 2019,
        month = apr,
       volume = {874},
       number = {2},
          eid = {L12},
        pages = {L12},
          doi = {10.3847/2041-8213/ab0e8c},
archivePrefix = {arXiv},
       eprint = {1901.03711},
 primaryClass = {astro-ph.GA},
       adsurl = {https://ui.adsabs.harvard.edu/abs/2019ApJ...874L..12D}
}

@ARTICLE{Danieli2020,
       author = {{Danieli}, Shany and {van Dokkum}, Pieter and {Abraham}, Roberto and {Conroy}, Charlie and {Dolphin}, Andrew E. and {Romanowsky}, Aaron J.},
        title = "{A Tip of the Red Giant Branch Distance to the Dark Matter Deficient Galaxy NGC 1052-DF4 from Deep Hubble Space Telescope Data}",
      journal = {\apjl},
         year = 2020,
        month = may,
       volume = {895},
       number = {1},
          eid = {L4},
        pages = {L4},
          doi = {10.3847/2041-8213/ab8dc4},
archivePrefix = {arXiv},
       eprint = {1910.07529},
 primaryClass = {astro-ph.GA},
       adsurl = {https://ui.adsabs.harvard.edu/abs/2020ApJ...895L...4D}
}

@software{Dolphin2016,
       author = {{Dolphin}, Andrew},
        title = "{DOLPHOT: Stellar photometry}",
 howpublished = {Astrophysics Source Code Library, record ascl:1608.013},
         year = 2016,
        month = aug,
          eid = {ascl:1608.013},
archivePrefix = {ascl},
       eprint = {1608.013},
       adsurl = {https://ui.adsabs.harvard.edu/abs/2016ascl.soft08013D}
}

@ARTICLE{Emsellem2019,
       author = {{Emsellem}, Eric and {van der Burg}, Remco F.~J. and {Fensch}, J{\'e}r{\'e}my and {Je{\v{r}}{\'a}bkov{\'a}}, Tereza and {Zanella}, Anita and {Agnello}, Adriano and {Hilker}, Michael and {M{\"u}ller}, Oliver and {Rejkuba}, Marina and {Duc}, Pierre-Alain and {Durrell}, Patrick and {Habas}, Rebecca and {Lelli}, Federico and {Lim}, Sungsoon and {Marleau}, Francine R. and {Peng}, Eric and {S{\'a}nchez-Janssen}, Rub{\'e}n},
        title = "{The ultra-diffuse galaxy NGC 1052-DF2 with MUSE. I. Kinematics of the stellar body}",
      journal = {\aap},
         year = 2019,
        month = may,
       volume = {625},
          eid = {A76},
        pages = {A76},
          doi = {10.1051/0004-6361/201834909},
archivePrefix = {arXiv},
       eprint = {1812.07345},
 primaryClass = {astro-ph.GA},
       adsurl = {https://ui.adsabs.harvard.edu/abs/2019A&A...625A..76E}
}

@ARTICLE{Fosbury1978,
       author = {{Fosbury}, R.~A.~E. and {Mebold}, U. and {Goss}, W.~M. and {Dopita}, M.~A.},
        title = "{The active elliptical galaxy NGC 1052.}",
      journal = {\mnras},
         year = 1978,
        month = jun,
       volume = {183},
        pages = {549-568},
          doi = {10.1093/mnras/183.4.549},
       adsurl = {https://ui.adsabs.harvard.edu/abs/1978MNRAS.183..549F}
}

@ARTICLE{Freedman2025,
       author = {{Freedman}, Wendy L. and {Madore}, Barry F. and {Hoyt}, Taylor J. and {Jang}, In Sung and {Lee}, Abigail J. and {Owens}, Kayla A.},
        title = "{Status Report on the Chicago-Carnegie Hubble Program (CCHP): Measurement of the Hubble Constant Using the Hubble and James Webb Space Telescopes}",
      journal = {\apj},
         year = 2025,
        month = jun,
       volume = {985},
       number = {2},
          eid = {203},
        pages = {203},
          doi = {10.3847/1538-4357/adce78},
archivePrefix = {arXiv},
       eprint = {2408.06153},
 primaryClass = {astro-ph.CO},
       adsurl = {https://ui.adsabs.harvard.edu/abs/2025ApJ...985..203F}
}

@ARTICLE{Forbes2019,
       author = {{Forbes}, Duncan A. and {Alabi}, Adebusola and {Brodie}, Jean P. and {Romanowsky}, Aaron J.},
        title = "{Dark matter and no dark matter: on the halo mass of NGC 1052}",
      journal = {\mnras},
         year = 2019,
        month = nov,
       volume = {489},
       number = {3},
        pages = {3665-3669},
          doi = {10.1093/mnras/stz2420},
archivePrefix = {arXiv},
       eprint = {1908.10858},
 primaryClass = {astro-ph.GA},
       adsurl = {https://ui.adsabs.harvard.edu/abs/2019MNRAS.489.3665F}
}

@BOOK{Fouque1992,
        author = {{Fouqu{\'e}}, P. and {Durand}, N. and {Bottinelli}, L. and {Gouguenheim}, L. and {Paturel}, G.},
        title={Catalogue of Optical Radial Velocities},
        isbn={9782908288056},
        series={Monographies de la base de donne{\'e}s extragalactiques},
        url={https://books.google.com/books?id=5luUAAAACAAJ},
        year={1992},
        publisher={Observatoire de Lyon}
}

@ARTICLE{Golini2024,
       author = {{Golini}, Giulia and {Montes}, Mireia and {Carrasco}, Eleazar R. and {Rom{\'a}n}, Javier and {Trujillo}, Ignacio},
        title = "{Ultra-deep imaging of NGC 1052-DF2 and NGC 1052-DF4 to unravel their origins}",
      journal = {\aap},
         year = 2024,
        month = apr,
       volume = {684},
          eid = {A99},
        pages = {A99},
          doi = {10.1051/0004-6361/202348300},
archivePrefix = {arXiv},
       eprint = {2402.04304},
 primaryClass = {astro-ph.GA},
       adsurl = {https://ui.adsabs.harvard.edu/abs/2024A&A...684A..99G}
}

@ARTICLE{Greco2021,
       author = {{Greco}, Johnny P. and {van Dokkum}, Pieter and {Danieli}, Shany and {Carlsten}, Scott G. and {Conroy}, Charlie},
        title = "{Measuring Distances to Low-luminosity Galaxies Using Surface Brightness Fluctuations}",
      journal = {\apj},
         year = 2021,
        month = feb,
       volume = {908},
       number = {1},
          eid = {24},
        pages = {24},
          doi = {10.3847/1538-4357/abd030},
archivePrefix = {arXiv},
       eprint = {2004.07273},
 primaryClass = {astro-ph.GA},
       adsurl = {https://ui.adsabs.harvard.edu/abs/2021ApJ...908...24G}
}

@ARTICLE{Jacoby2024,
       author = {{Jacoby}, George H. and {Ciardullo}, Robin and {Roth}, Martin M. and {Arnaboldi}, Magda and {Weilbacher}, Peter M.},
        title = "{Toward Precision Cosmology with Improved Planetary Nebula Luminosity Function Distances Using VLT-MUSE. II. A Test Sample from Archival Data}",
      journal = {\apjs},
         year = 2024,
        month = apr,
       volume = {271},
       number = {2},
          eid = {40},
        pages = {40},
          doi = {10.3847/1538-4365/ad2166},
archivePrefix = {arXiv},
       eprint = {2309.11603},
 primaryClass = {astro-ph.CO},
       adsurl = {https://ui.adsabs.harvard.edu/abs/2024ApJS..271...40J}
}

@ARTICLE{Jang2017,
       author = {{Jang}, In Sung and {Lee}, Myung Gyoon},
        title = "{The Tip of the Red Giant Branch Distances to Typa Ia Supernova Host Galaxies. V. NGC 3021, NGC 3370, and NGC 1309 and the Value of the Hubble Constant}",
      journal = {\apj},
         year = 2017,
        month = feb,
       volume = {836},
       number = {1},
          eid = {74},
        pages = {74},
          doi = {10.3847/1538-4357/836/1/74},
       adsurl = {https://ui.adsabs.harvard.edu/abs/2017ApJ...836...74J}
}

@ARTICLE{Karachentsev2000,
       author = {{Karachentsev}, I.~D. and {Karachentseva}, V.~E. and {Suchkov}, A.~A. and {Grebel}, E.~K.},
        title = "{Dwarf galaxy candidates found on the SERC EJ sky survey}",
      journal = {\aaps},
         year = 2000,
        month = sep,
       volume = {145},
        pages = {415-423},
          doi = {10.1051/aas:2000249},
       adsurl = {https://ui.adsabs.harvard.edu/abs/2000A&AS..145..415K}
}

@ARTICLE{Keim2022,
       author = {{Keim}, Michael A. and {van Dokkum}, Pieter and {Danieli}, Shany and {Lokhorst}, Deborah and {Li}, Jiaxuan and {Shen}, Zili and {Abraham}, Roberto and {Chen}, Seery and {Gilhuly}, Colleen and {Liu}, Qing and {Merritt}, Allison and {Miller}, Tim B. and {Pasha}, Imad and {Polzin}, Ava},
        title = "{Tidal Distortions in NGC1052-DF2 and NGC1052-DF4: Independent Evidence for a Lack of Dark Matter}",
      journal = {\apj},
         year = 2022,
        month = aug,
       volume = {935},
       number = {2},
          eid = {160},
        pages = {160},
          doi = {10.3847/1538-4357/ac7dab},
archivePrefix = {arXiv},
       eprint = {2109.09778},
 primaryClass = {astro-ph.GA},
       adsurl = {https://ui.adsabs.harvard.edu/abs/2022ApJ...935..160K}
}

@ARTICLE{Keim2025,
       author = {{Keim}, Michael A. and {van Dokkum}, Pieter and {Shen}, Zili and {Souchereau}, Harrison and {Pasha}, Imad and {Danieli}, Shany and {Abraham}, Roberto and {Romanowsky}, Aaron J. and {Tang}, Yimeng},
        title = "{Kinematic Confirmation of a Remarkable Linear Trail of Galaxies in the NGC 1052 Field, Consistent with Formation in a High-speed Bullet Dwarf Collision}",
      journal = {\apj},
         year = 2025,
        month = aug,
       volume = {988},
       number = {2},
          eid = {165},
        pages = {165},
          doi = {10.3847/1538-4357/addfd4},
archivePrefix = {arXiv},
       eprint = {2506.10220},
 primaryClass = {astro-ph.GA},
       adsurl = {https://ui.adsabs.harvard.edu/abs/2025ApJ...988..165K}
}

@misc{Keim2026,
      title={A Third Galaxy Missing Dark Matter along a Trail of Galaxies in the NGC 1052 Field}, 
      author={Michael A. Keim and Pieter van Dokkum and Zili Shen and Shany Danieli and Imad Pasha},
      year={2026},
      eprint={2603.15860},
      archivePrefix={arXiv},
      primaryClass={astro-ph.GA},
      url={https://arxiv.org/abs/2603.15860}, 
}

@ARTICLE{Kim2021,
       author = {{Kim}, Yoo Jung and {Lee}, Myung Gyoon},
        title = "{Calibration of Surface Brightness Fluctuations for Dwarf Galaxies in the Hyper Suprime-Cam gi Filter System}",
      journal = {\apj},
         year = 2021,
        month = dec,
       volume = {923},
       number = {2},
          eid = {152},
        pages = {152},
          doi = {10.3847/1538-4357/ac2d94},
archivePrefix = {arXiv},
       eprint = {2110.02522},
 primaryClass = {astro-ph.GA},
       adsurl = {https://ui.adsabs.harvard.edu/abs/2021ApJ...923..152K}
}

@ARTICLE{Lee1993,
       author = {{Lee}, Myung Gyoon and {Freedman}, Wendy L. and {Madore}, Barry F.},
        title = "{The Tip of the Red Giant Branch as a Distance Indicator for Resolved Galaxies}",
      journal = {\apj},
         year = 1993,
        month = nov,
       volume = {417},
        pages = {553},
          doi = {10.1086/173334},
       adsurl = {https://ui.adsabs.harvard.edu/abs/1993ApJ...417..553L}
}

@ARTICLE{Lee2021,
       author = {{Lee}, Joohyun and {Shin}, Eun-jin and {Kim}, Ji-hoon},
        title = "{Dark Matter Deficient Galaxies and Their Member Star Clusters Form Simultaneously during High-velocity Galaxy Collisions in 1.25 pc Resolution Simulations}",
      journal = {\apjl},
         year = 2021,
        month = aug,
       volume = {917},
       number = {2},
          eid = {L15},
        pages = {L15},
          doi = {10.3847/2041-8213/ac16e0},
archivePrefix = {arXiv},
       eprint = {2108.01102},
 primaryClass = {astro-ph.GA},
       adsurl = {https://ui.adsabs.harvard.edu/abs/2021ApJ...917L..15L}
}

@ARTICLE{Lee2024,
       author = {{Lee}, Joohyun and {Shin}, Eun-jin and {Kim}, Ji-hoon and {Shapiro}, Paul R. and {Chung}, Eunwoo},
        title = "{Multiple Beads on a String: Dark-matter-deficient Galaxy Formation in a Mini-Bullet Satellite{\textendash}Satellite Galaxy Collision}",
      journal = {\apj},
         year = 2024,
        month = may,
       volume = {966},
       number = {1},
          eid = {72},
        pages = {72},
          doi = {10.3847/1538-4357/ad2932},
archivePrefix = {arXiv},
       eprint = {2312.11350},
 primaryClass = {astro-ph.GA},
       adsurl = {https://ui.adsabs.harvard.edu/abs/2024ApJ...966...72L}
}

@ARTICLE{Li2024,
       author = {{Li}, Jiaxuan and {Greene}, Jenny E. and {Carlsten}, Scott G. and {Danieli}, Shany},
        title = "{Hedgehog: An Isolated Quiescent Dwarf Galaxy at 2.4 Mpc}",
      journal = {\apjl},
         year = 2024,
        month = nov,
       volume = {975},
       number = {1},
          eid = {L23},
        pages = {L23},
          doi = {10.3847/2041-8213/ad5b59},
archivePrefix = {arXiv},
       eprint = {2406.00101},
 primaryClass = {astro-ph.GA},
       adsurl = {https://ui.adsabs.harvard.edu/abs/2024ApJ...975L..23L}
}

@ARTICLE{Madore2023,
       author = {{Madore}, Barry F. and {Freedman}, Wendy L. and {Owens}, Kayla A. and {Jang}, In Sung},
        title = "{Quantifying Uncertainties on the Tip of the Red Giant Branch Method}",
      journal = {\aj},
         year = 2023,
        month = jul,
       volume = {166},
       number = {1},
          eid = {2},
        pages = {2},
          doi = {10.3847/1538-3881/acd3f3},
archivePrefix = {arXiv},
       eprint = {2305.06195},
 primaryClass = {astro-ph.SR},
       adsurl = {https://ui.adsabs.harvard.edu/abs/2023AJ....166....2M}
}

@ARTICLE{Makarov2006,
       author = {{Makarov}, Dmitry and {Makarova}, Lidia and {Rizzi}, Luca and {Tully}, R. Brent and {Dolphin}, Andrew E. and {Sakai}, Shoko and {Shaya}, Edward J.},
        title = "{Tip of the Red Giant Branch Distances. I. Optimization of a Maximum Likelihood Algorithm}",
      journal = {\aj},
         year = 2006,
        month = dec,
       volume = {132},
       number = {6},
        pages = {2729-2742},
          doi = {10.1086/508925},
archivePrefix = {arXiv},
       eprint = {astro-ph/0603073},
 primaryClass = {astro-ph},
       adsurl = {https://ui.adsabs.harvard.edu/abs/2006AJ....132.2729M}
}

@ARTICLE{Mei2005,
       author = {{Mei}, Simona and {Blakeslee}, John P. and {Tonry}, John L. and {Jord{\'a}n}, Andr{\'e}s and {Peng}, Eric W. and {C{\^o}t{\'e}}, Patrick and {Ferrarese}, Laura and {Merritt}, David and {Milosavljevi{\'c}}, Milo{\v{s}} and {West}, Michael J.},
        title = "{The ACS Virgo Cluster Survey. IV. Data Reduction Procedures for Surface Brightness Fluctuation Measurements with the Advanced Camera for Surveys}",
      journal = {\apjs},
         year = 2005,
        month = feb,
       volume = {156},
       number = {2},
        pages = {113-125},
          doi = {10.1086/426544},
archivePrefix = {arXiv},
       eprint = {astro-ph/0501325},
 primaryClass = {astro-ph},
       adsurl = {https://ui.adsabs.harvard.edu/abs/2005ApJS..156..113M}
}

@ARTICLE{Monelli2019,
       author = {{Monelli}, Matteo and {Trujillo}, Ignacio},
        title = "{The TRGB Distance to the Second Galaxy {\textquotedblleft}Missing Dark Matter{\textquotedblright}: Evidence for Two Groups of Galaxies at 13.5 and 19 Mpc in the Line of Sight of NGC 1052}",
      journal = {\apjl},
         year = 2019,
        month = jul,
       volume = {880},
       number = {1},
          eid = {L11},
        pages = {L11},
          doi = {10.3847/2041-8213/ab2fd2},
archivePrefix = {arXiv},
       eprint = {1907.03761},
 primaryClass = {astro-ph.GA},
       adsurl = {https://ui.adsabs.harvard.edu/abs/2019ApJ...880L..11M}
}

@ARTICLE{Montes2020,
       author = {{Montes}, Mireia and {Infante-Sainz}, Ra{\'u}l and {Madrigal-Aguado}, Alberto and {Rom{\'a}n}, Javier and {Monelli}, Matteo and {Borlaff}, Alejandro S. and {Trujillo}, Ignacio},
        title = "{The Galaxy ``Missing Dark Matter'' NGC 1052-DF4 is Undergoing Tidal Disruption}",
      journal = {\apj},
         year = 2020,
        month = dec,
       volume = {904},
       number = {2},
          eid = {114},
        pages = {114},
          doi = {10.3847/1538-4357/abc340},
archivePrefix = {arXiv},
       eprint = {2010.09719},
 primaryClass = {astro-ph.GA},
       adsurl = {https://ui.adsabs.harvard.edu/abs/2020ApJ...904..114M}
}

@MISC{Morishita2023,
       author = {{Morishita}, Takahiro and {Mason}, Charlotte and {Trenti}, Michele and {Treu}, Tommaso L. and {Abdurro'uf}, Abdurro'uf and {Alavi}, Anahita and {Atek}, Hakim and {Bahe}, Yannick and {Bradac}, Marusa and {Bradley}, Larry and {Bunker}, Andrew and {Coe}, Dan and {Colbert}, James and {Hayes}, Matthew James and {Jones}, Tucker and {Kodama}, Tadayuki and {Leethochawalit}, Nicha and {Liu}, Zhaoran and {Malkan}, Matthew A. and {Mehta}, Vihang and {Metha}, Benjamin Andrew and {Newman}, Andrew B. and {Rafelski}, Marc and {Roberts-Borsani}, Guido and {Rutkowski}, Michael James and {Scarlata}, Claudia and {Stiavelli}, Massimo and {Teplitz}, Harry and {Vulcani}, Benedetta and {Wang}, Xin},
        title = "{A NIRCam Pure-Parallel Imaging Survey of Galaxies Across the Universe}",
 howpublished = {JWST Proposal. Cycle 2, ID. \#3990},
         year = 2023,
        month = may,
        pages = {3990},
       adsurl = {https://ui.adsabs.harvard.edu/abs/2023jwst.prop.3990M}
}

@ARTICLE{Newman2024_hst,
       author = {{Newman}, Max J.~B. and {McQuinn}, Kristen B.~W. and {Skillman}, Evan D. and {Boyer}, Martha L. and {Cohen}, Roger E. and {Dolphin}, Andrew E. and {Telford}, O. Grace},
        title = "{An Empirical Calibration of the Tip of the Red Giant Branch Distance Method in the Near Infrared. I. Hubble Space Telescope WFC3/IR F110W and F160W Filters}",
      journal = {\apj},
         year = 2024,
        month = may,
       volume = {966},
       number = {2},
          eid = {175},
        pages = {175},
          doi = {10.3847/1538-4357/ad306d},
archivePrefix = {arXiv},
       eprint = {2403.03086},
 primaryClass = {astro-ph.GA},
       adsurl = {https://ui.adsabs.harvard.edu/abs/2024ApJ...966..175N}
}

@ARTICLE{Newman2024_jwst,
       author = {{Newman}, Max J.~B. and {McQuinn}, Kristen B.~W. and {Skillman}, Evan D. and {Boyer}, Martha L. and {Cohen}, Roger E. and {Dolphin}, Andrew E. and {Telford}, O. Grace},
        title = "{An Empirical Calibration of the Tip of the Red Giant Branch Distance Method in the Near Infrared. II. JWST NIRCam Wide Filters}",
      journal = {\apj},
         year = 2024,
        month = nov,
       volume = {975},
       number = {2},
          eid = {195},
        pages = {195},
          doi = {10.3847/1538-4357/ad79f8},
archivePrefix = {arXiv},
       eprint = {2406.03532},
 primaryClass = {astro-ph.CO},
       adsurl = {https://ui.adsabs.harvard.edu/abs/2024ApJ...975..195N}
}

@INPROCEEDINGS{Otaki2022,
       author = {{Otaki}, Koki and {Mori}, Masao},
        title = "{The formation of dark-matter-deficient galaxies through galaxy collisions}",
    booktitle = {Journal of Physics Conference Series},
         year = 2022,
       series = {Journal of Physics Conference Series},
       volume = {2207},
        month = mar,
    publisher = {IOP},
          eid = {012049},
        pages = {012049},
          doi = {10.1088/1742-6596/2207/1/012049},
       adsurl = {https://ui.adsabs.harvard.edu/abs/2022JPhCS2207a2049O}
}

@ARTICLE{Peng2002,
       author = {{Peng}, Chien Y. and {Ho}, Luis C. and {Impey}, Chris D. and {Rix}, Hans-Walter},
        title = "{Detailed Structural Decomposition of Galaxy Images}",
      journal = {\aj},
         year = 2002,
        month = jul,
       volume = {124},
       number = {1},
        pages = {266-293},
          doi = {10.1086/340952},
archivePrefix = {arXiv},
       eprint = {astro-ph/0204182},
 primaryClass = {astro-ph},
       adsurl = {https://ui.adsabs.harvard.edu/abs/2002AJ....124..266P}
}

@ARTICLE{Schlafly2011,
       author = {{Schlafly}, Edward F. and {Finkbeiner}, Douglas P.},
        title = "{Measuring Reddening with Sloan Digital Sky Survey Stellar Spectra and Recalibrating SFD}",
      journal = {\apj},
         year = 2011,
        month = aug,
       volume = {737},
       number = {2},
          eid = {103},
        pages = {103},
          doi = {10.1088/0004-637X/737/2/103},
archivePrefix = {arXiv},
       eprint = {1012.4804},
 primaryClass = {astro-ph.GA},
       adsurl = {https://ui.adsabs.harvard.edu/abs/2011ApJ...737..103S}
}

@ARTICLE{Shen2021a,
       author = {{Shen}, Zili and {van Dokkum}, Pieter and {Danieli}, Shany},
        title = "{A Complex Luminosity Function for the Anomalous Globular Clusters in NGC 1052-DF2 and NGC 1052-DF4}",
      journal = {\apj},
         year = 2021,
        month = mar,
       volume = {909},
       number = {2},
          eid = {179},
        pages = {179},
          doi = {10.3847/1538-4357/abdd29},
archivePrefix = {arXiv},
       eprint = {2010.07324},
 primaryClass = {astro-ph.GA},
       adsurl = {https://ui.adsabs.harvard.edu/abs/2021ApJ...909..179S}
}

@ARTICLE{Shen2021b,
       author = {{Shen}, Zili and {Danieli}, Shany and {van Dokkum}, Pieter and {Abraham}, Roberto and {Brodie}, Jean P. and {Conroy}, Charlie and {Dolphin}, Andrew E. and {Romanowsky}, Aaron J. and {Kruijssen}, J.~M. Diederik and {Dutta Chowdhury}, Dhruba},
        title = "{A Tip of the Red Giant Branch Distance of 22.1 {\ensuremath{\pm}} 1.2 Mpc to the Dark Matter Deficient Galaxy NGC 1052-DF2 from 40 Orbits of Hubble Space Telescope Imaging}",
      journal = {\apjl},
         year = 2021,
        month = jun,
       volume = {914},
       number = {1},
          eid = {L12},
        pages = {L12},
          doi = {10.3847/2041-8213/ac0335},
archivePrefix = {arXiv},
       eprint = {2104.03319},
 primaryClass = {astro-ph.GA},
       adsurl = {https://ui.adsabs.harvard.edu/abs/2021ApJ...914L..12S}
}

@ARTICLE{Shen2023,
       author = {{Shen}, Zili and {van Dokkum}, Pieter and {Danieli}, Shany},
        title = "{Confirmation of an Anomalously Low Dark Matter Content for the Galaxy NGC 1052-DF4 from Deep, High-resolution Continuum Spectroscopy}",
      journal = {\apj},
         year = 2023,
        month = nov,
       volume = {957},
       number = {1},
          eid = {6},
        pages = {6},
          doi = {10.3847/1538-4357/acfa70},
archivePrefix = {arXiv},
       eprint = {2309.08592},
 primaryClass = {astro-ph.GA},
       adsurl = {https://ui.adsabs.harvard.edu/abs/2023ApJ...957....6S}
}

@ARTICLE{Shin2020,
       author = {{Shin}, Eun-jin and {Jung}, Minyong and {Kwon}, Goojin and {Kim}, Ji-hoon and {Lee}, Joohyun and {Jo}, Yongseok and {Oh}, Boon Kiat},
        title = "{Dark Matter Deficient Galaxies Produced via High-velocity Galaxy Collisions in High-resolution Numerical Simulations}",
      journal = {\apj},
         year = 2020,
        month = aug,
       volume = {899},
       number = {1},
          eid = {25},
        pages = {25},
          doi = {10.3847/1538-4357/aba434},
archivePrefix = {arXiv},
       eprint = {2007.09889},
 primaryClass = {astro-ph.GA},
       adsurl = {https://ui.adsabs.harvard.edu/abs/2020ApJ...899...25S}
}

@ARTICLE{Silk2019,
       author = {{Silk}, Joseph},
        title = "{Ultra-diffuse galaxies without dark matter}",
      journal = {\mnras},
         year = 2019,
        month = sep,
       volume = {488},
       number = {1},
        pages = {L24-L28},
          doi = {10.1093/mnrasl/slz090},
archivePrefix = {arXiv},
       eprint = {1905.13235},
 primaryClass = {astro-ph.GA},
       adsurl = {https://ui.adsabs.harvard.edu/abs/2019MNRAS.488L..24S}
}

@ARTICLE{Tang2025a,
       author = {{Tang}, Yimeng and {Romanowsky}, Aaron J. and {van Dokkum}, Pieter G. and {Jarrett}, T.~H. and {Bundy}, Kevin A. and {Buzzo}, Maria Luisa and {Danieli}, Shany and {Gannon}, Jonah S. and {Keim}, Michael A. and {Laine}, Seppo and {Shen}, Zili},
        title = "{Testing the Bullet Dwarf Collision Scenario in the NGC 1052 Group through Morphologies and Stellar Populations}",
      journal = {\apj},
         year = 2025,
        month = jan,
       volume = {978},
       number = {1},
          eid = {21},
        pages = {21},
          doi = {10.3847/1538-4357/ad8cd0},
archivePrefix = {arXiv},
       eprint = {2410.19331},
 primaryClass = {astro-ph.GA},
       adsurl = {https://ui.adsabs.harvard.edu/abs/2025ApJ...978...21T}
}

@ARTICLE{Tang2025b,
       author = {{Tang}, Yimeng and {Romanowsky}, Aaron J. and {Gannon}, Jonah S. and {Janssens}, Steven R. and {Brodie}, Jean P. and {Bundy}, Kevin A. and {Buzzo}, Maria Luisa and {Cabrera}, Enrique A. and {Danieli}, Shany and {Ferr{\'e}-Mateu}, Anna and {Forbes}, Duncan A. and {van Dokkum}, Pieter G.},
        title = "{An Unexplained Origin for the Unusual Globular Cluster System in the Ultradiffuse Galaxy FCC 224}",
      journal = {\apj},
         year = 2025,
        month = mar,
       volume = {982},
       number = {1},
          eid = {1},
        pages = {1},
          doi = {10.3847/1538-4357/adae11},
archivePrefix = {arXiv},
       eprint = {2501.10665},
 primaryClass = {astro-ph.GA},
       adsurl = {https://ui.adsabs.harvard.edu/abs/2025ApJ...982....1T}
}

@ARTICLE{Tonry1988,
       author = {{Tonry}, John and {Schneider}, Donald P.},
        title = "{A New Technique for Measuring Extragalactic Distances}",
      journal = {\aj},
         year = 1988,
        month = sep,
       volume = {96},
        pages = {807},
          doi = {10.1086/114847},
       adsurl = {https://ui.adsabs.harvard.edu/abs/1988AJ.....96..807T}
}

@ARTICLE{Trujillo2019,
       author = {{Trujillo}, Ignacio and {Beasley}, Michael A. and {Borlaff}, Alejandro and {Carrasco}, Eleazar R. and {Di Cintio}, Arianna and {Filho}, Mercedes and {Monelli}, Matteo and {Montes}, Mireia and {Rom{\'a}n}, Javier and {Ruiz-Lara}, Tom{\'a}s and {S{\'a}nchez Almeida}, Jorge and {Valls-Gabaud}, David and {Vazdekis}, Alexandre},
        title = "{A distance of 13 Mpc resolves the claimed anomalies of the galaxy lacking dark matter}",
      journal = {\mnras},
         year = 2019,
        month = jun,
       volume = {486},
       number = {1},
        pages = {1192-1219},
          doi = {10.1093/mnras/stz771},
archivePrefix = {arXiv},
       eprint = {1806.10141},
 primaryClass = {astro-ph.GA},
       adsurl = {https://ui.adsabs.harvard.edu/abs/2019MNRAS.486.1192T}
}

@ARTICLE{vanDokkum2018a,
       author = {{van Dokkum}, Pieter and {Danieli}, Shany and {Cohen}, Yotam and {Merritt}, Allison and {Romanowsky}, Aaron J. and {Abraham}, Roberto and {Brodie}, Jean and {Conroy}, Charlie and {Lokhorst}, Deborah and {Mowla}, Lamiya and {O'Sullivan}, Ewan and {Zhang}, Jielai},
        title = "{A galaxy lacking dark matter}",
      journal = {\nat},
         year = 2018,
        month = mar,
       volume = {555},
       number = {7698},
        pages = {629-632},
          doi = {10.1038/nature25767},
archivePrefix = {arXiv},
       eprint = {1803.10237},
 primaryClass = {astro-ph.GA},
       adsurl = {https://ui.adsabs.harvard.edu/abs/2018Natur.555..629V}
}

@ARTICLE{vanDokkum2018b,
       author = {{van Dokkum}, Pieter and {Cohen}, Yotam and {Danieli}, Shany and {Kruijssen}, J.~M. Diederik and {Romanowsky}, Aaron J. and {Merritt}, Allison and {Abraham}, Roberto and {Brodie}, Jean and {Conroy}, Charlie and {Lokhorst}, Deborah and {Mowla}, Lamiya and {O'Sullivan}, Ewan and {Zhang}, Jielai},
        title = "{An Enigmatic Population of Luminous Globular Clusters in a Galaxy Lacking Dark Matter}",
      journal = {\apjl},
         year = 2018,
        month = apr,
       volume = {856},
       number = {2},
          eid = {L30},
        pages = {L30},
          doi = {10.3847/2041-8213/aab60b},
archivePrefix = {arXiv},
       eprint = {1803.10240},
 primaryClass = {astro-ph.GA},
       adsurl = {https://ui.adsabs.harvard.edu/abs/2018ApJ...856L..30V}
}

@ARTICLE{vanDokkum2018d,
       author = {{van Dokkum}, Pieter and {Danieli}, Shany and {Cohen}, Yotam and {Romanowsky}, Aaron J. and {Conroy}, Charlie},
        title = "{The Distance of the Dark Matter Deficient Galaxy NGC 1052-DF2}",
      journal = {\apjl},
         year = 2018,
        month = sep,
       volume = {864},
       number = {1},
          eid = {L18},
        pages = {L18},
          doi = {10.3847/2041-8213/aada4d},
archivePrefix = {arXiv},
       eprint = {1807.06025},
 primaryClass = {astro-ph.GA},
       adsurl = {https://ui.adsabs.harvard.edu/abs/2018ApJ...864L..18V}
}
\bibliographystyle{aasjournal}

\restartappendixnumbering
\appendix

\section{SBF fitting for all galaxies}
\label{sbf_all}

Figure \ref{fig:sbf_all} shows the SBF fitting for all galaxies in our sample, in addition to DF2 and DF7, which are presented in Figure \ref{fig:sbf}.

\begin{figure*}
\centering
\includegraphics[width=0.9\textwidth]{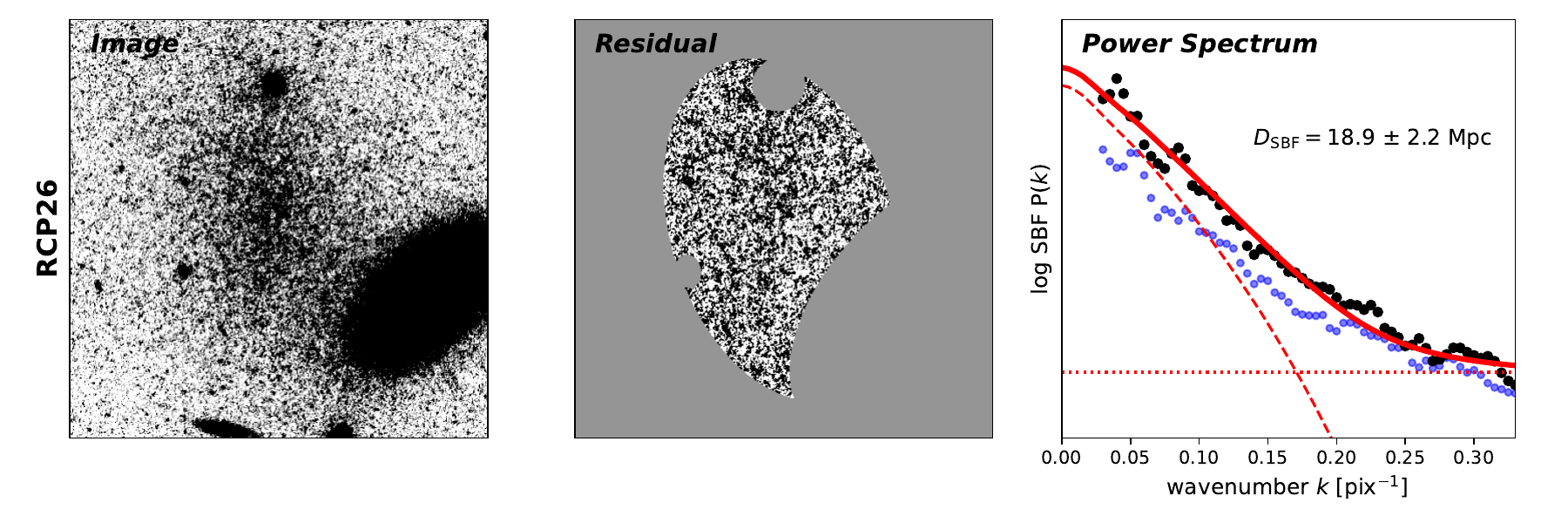}
\includegraphics[width=0.9\textwidth]{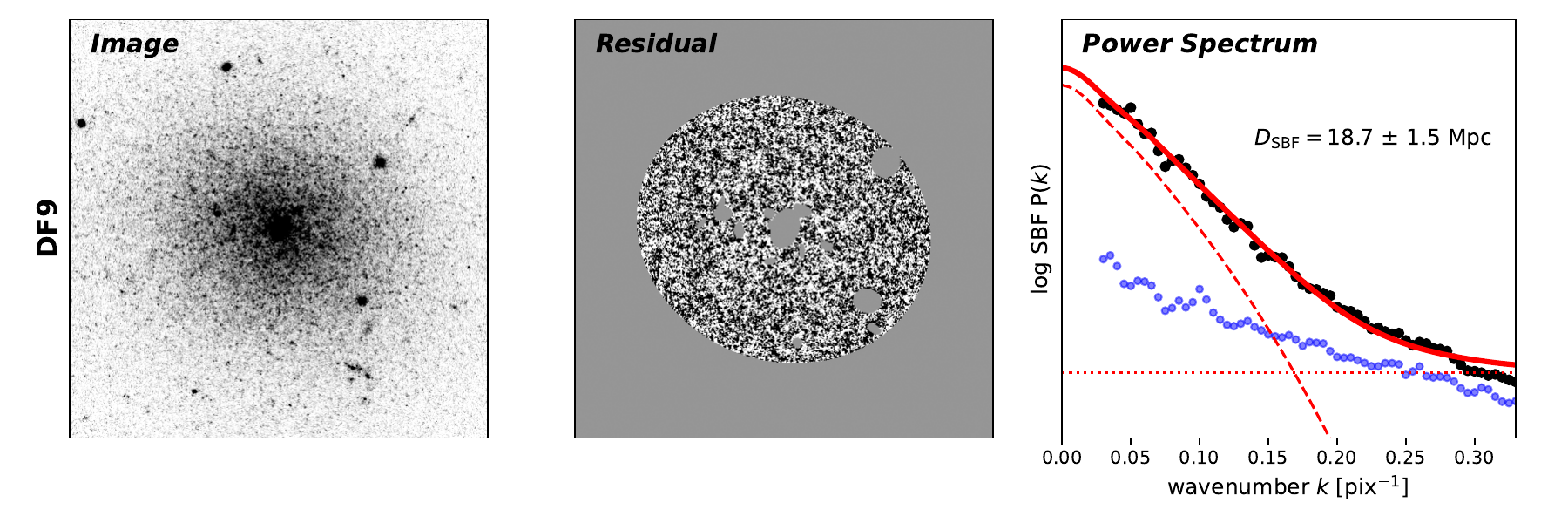}
\includegraphics[width=0.9\textwidth]{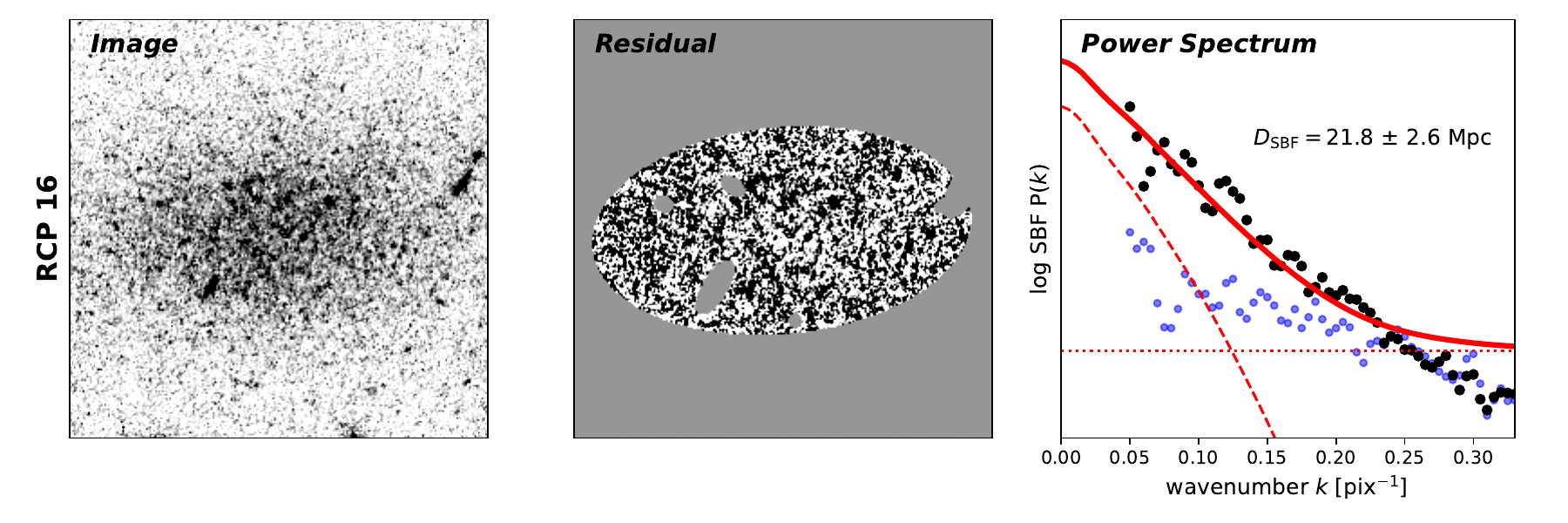}
\includegraphics[width=0.9\textwidth]{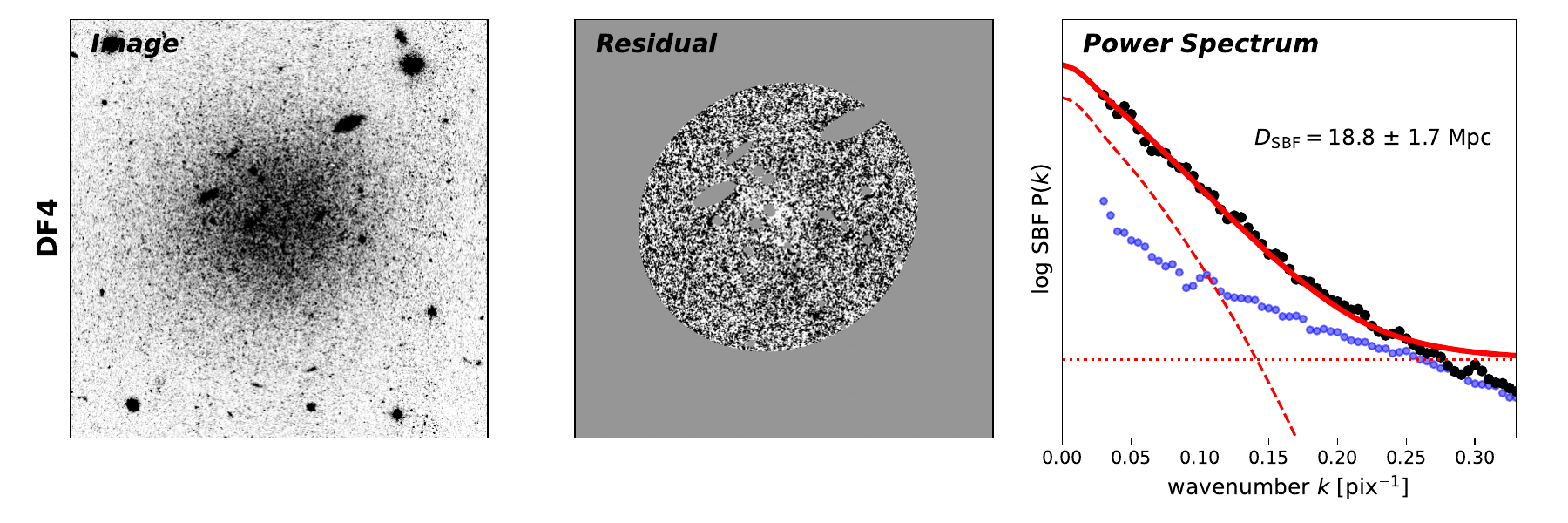}
\caption{As Figure \ref{fig:sbf}, but for other galaxies.}
\label{fig:sbf_all}
\end{figure*}

\begin{figure*}
\ContinuedFloat
\centering
\includegraphics[width=0.9\textwidth]{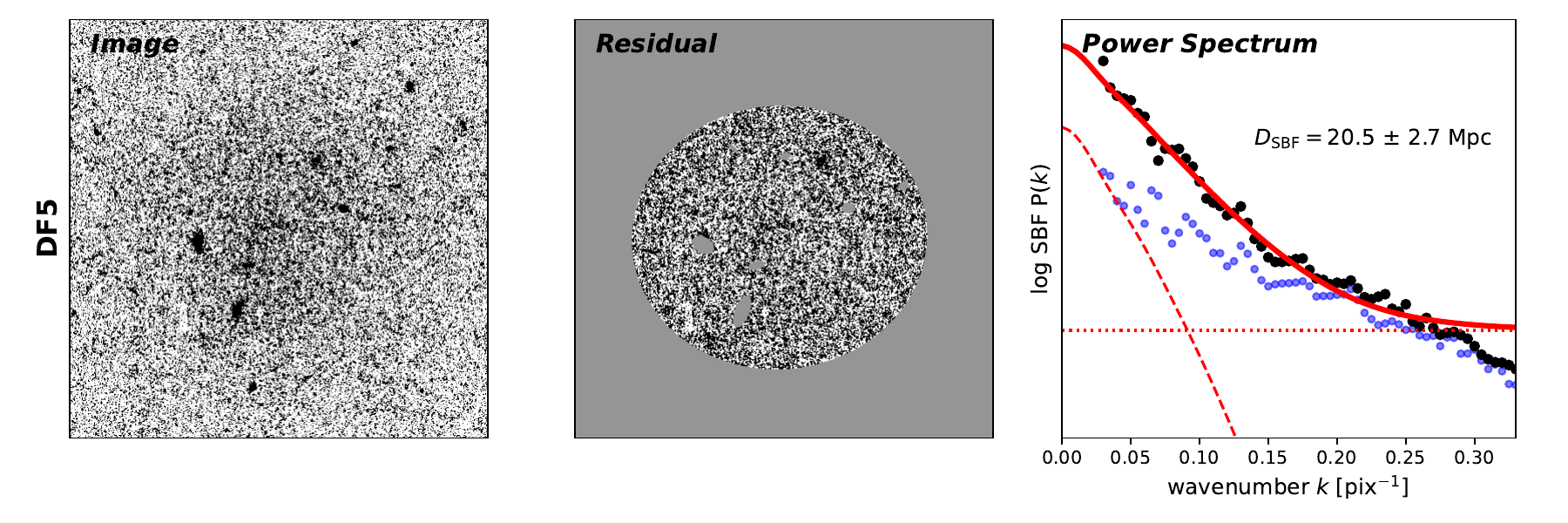}
\includegraphics[width=0.9\textwidth]{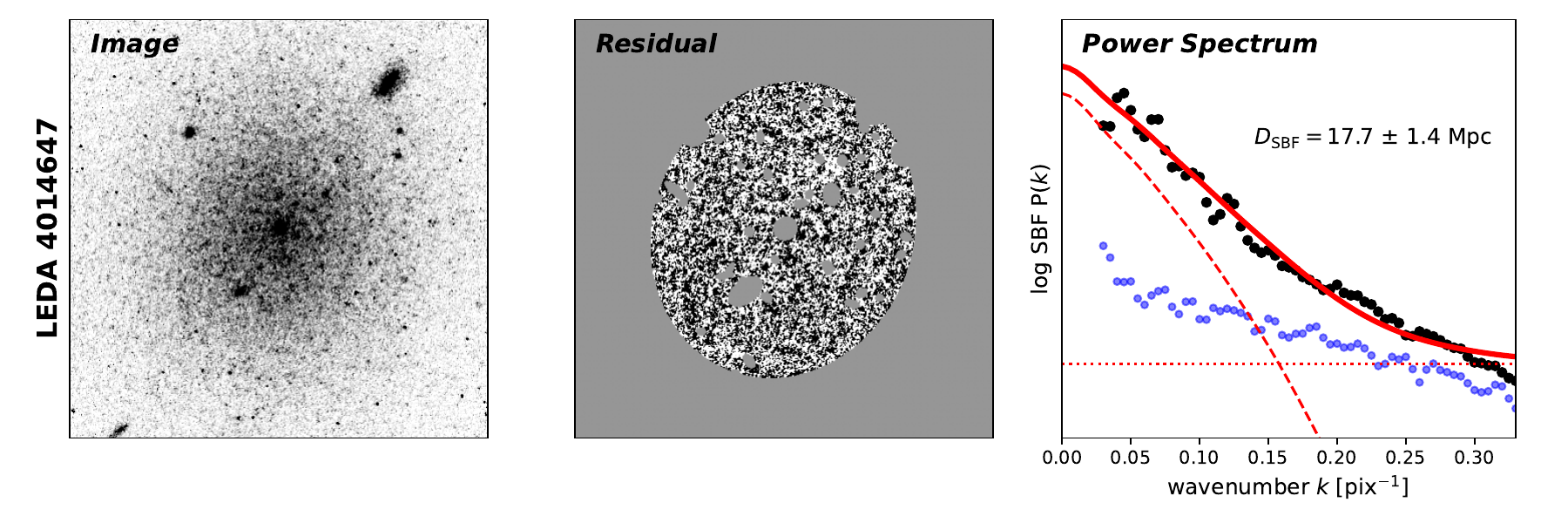}
\includegraphics[width=0.9\textwidth]{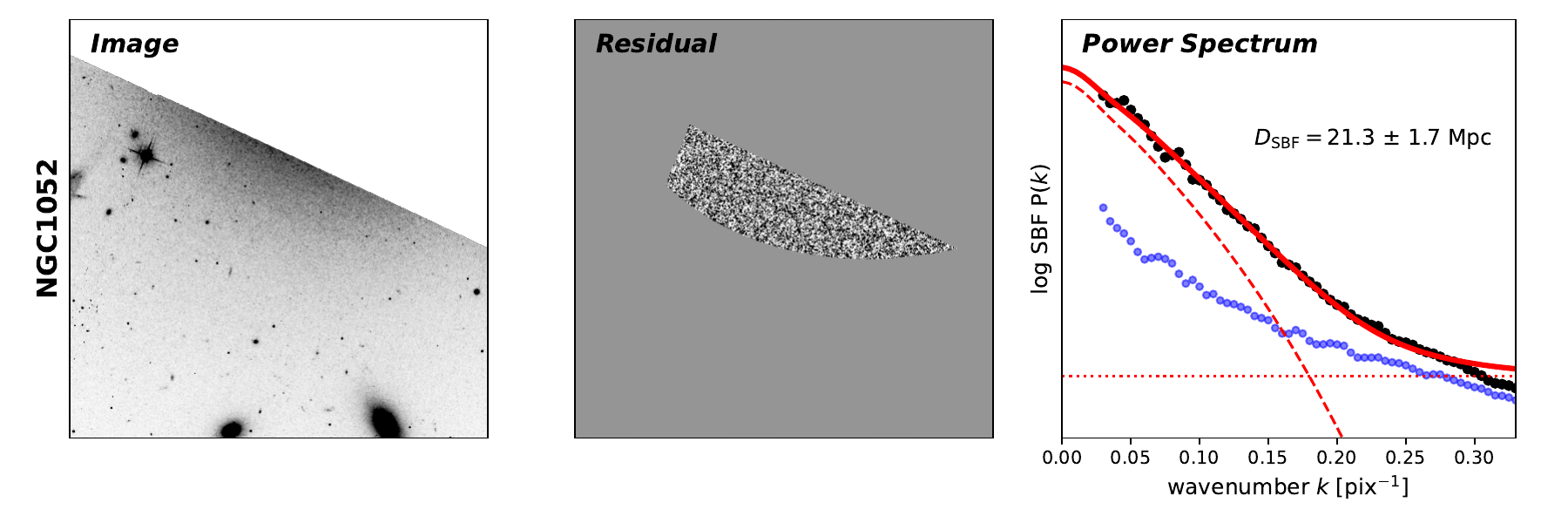}
\includegraphics[width=0.9\textwidth]{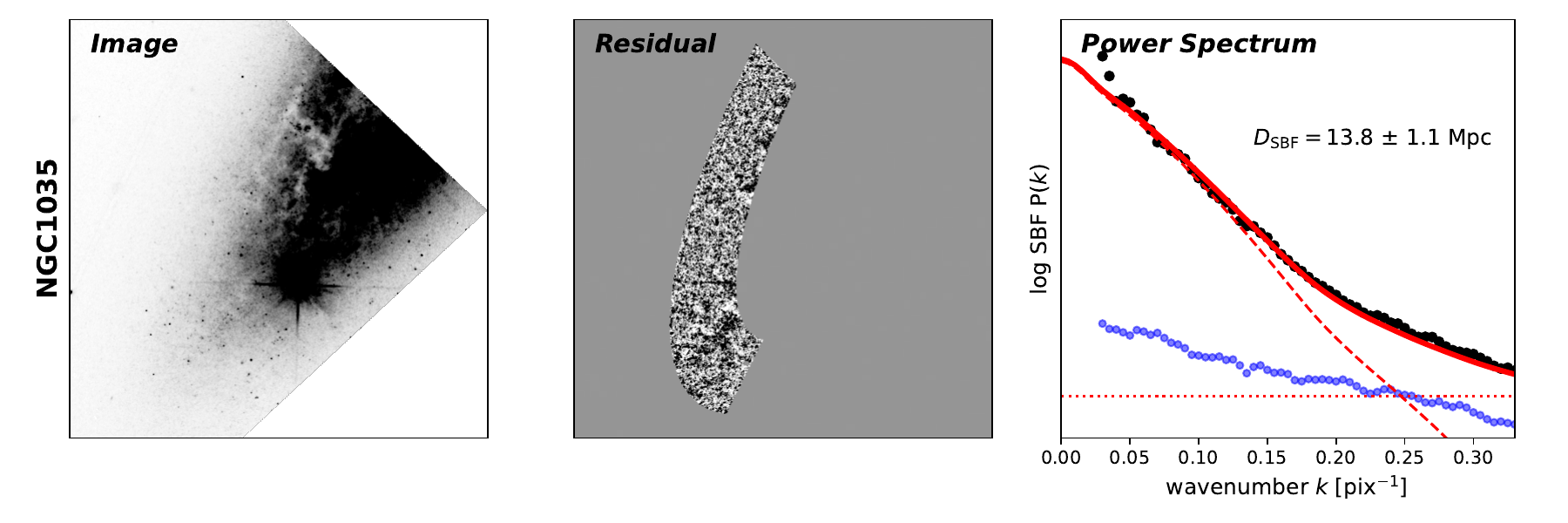}
\caption{Continued. As Figure \ref{fig:sbf}, but for other galaxies. NGC 1052 and NGC 1035 are exceptions in terms of their masked regions shown in the middle panels, which are defined by image boundaries and isophotal ellipses.}
\end{figure*}

\section{JWST Artificial Star Tests}
\label{art_stars}

\begin{figure*}
\centering
\includegraphics[width=0.95\textwidth]{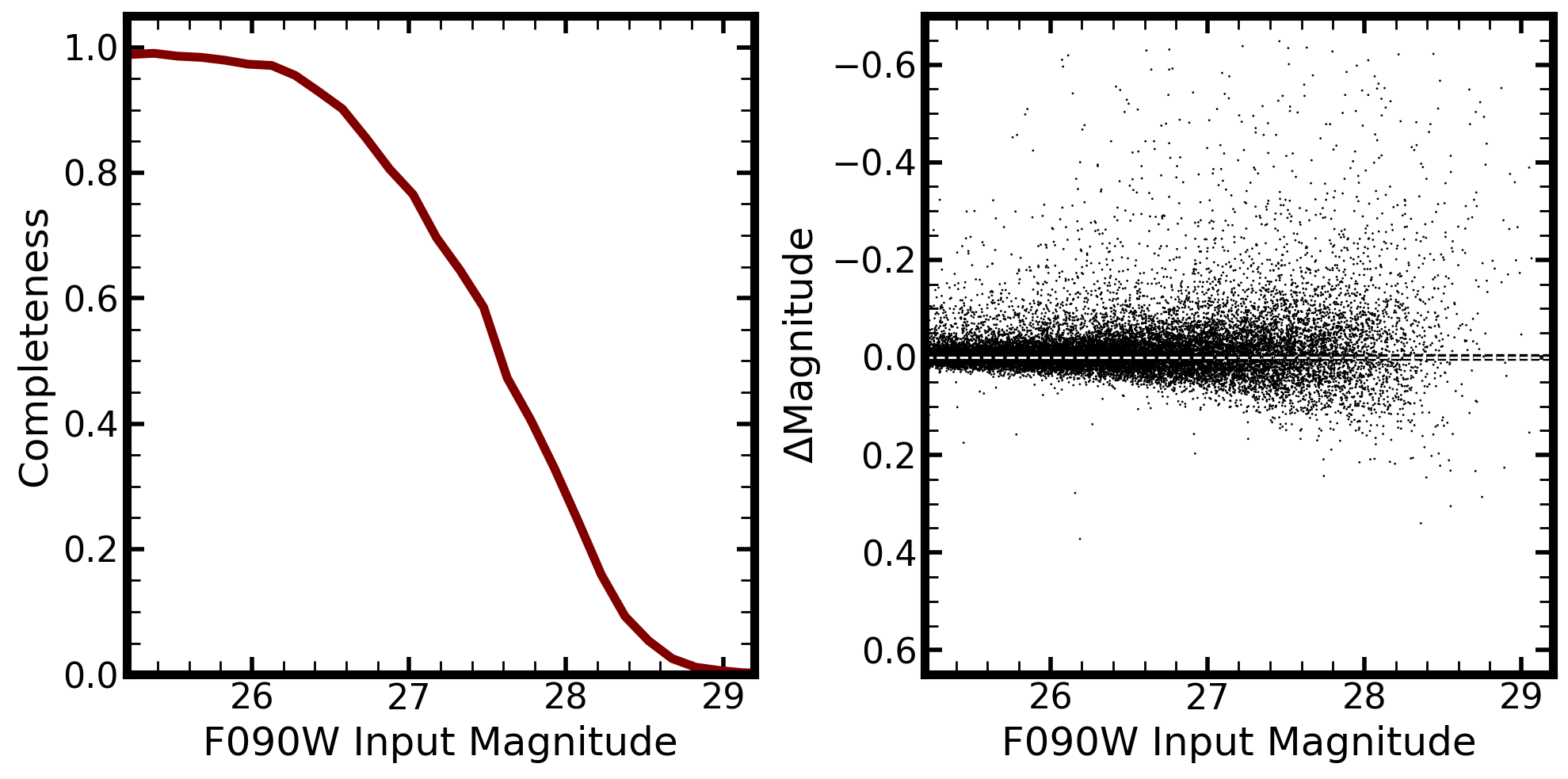}
\caption{\textbf{Left:} Photometric completeness as a function of F090W input magnitude. \textbf{Right:} Difference between output and input artificial stellar magnitudes as a function of F090W input magnitude.}
\label{fig:jwst_art_stars}
\end{figure*}

We performed artificial star tests with \texttt{DOLPHOT} to quantify the photometric bias, completeness, and errors present in the archival JWST data. Using the same input parameters as our genuine photometry, we injected and attempted to recover $\sim$100,000 artificial stars. This procedure took place one star at a time, so as to not artificially increase the level of stellar crowding present in the images. Summary plots are shown in Figure \ref{fig:jwst_art_stars} as a function of input F090W magnitude, where the range of input colors is limited to those seen along the RGB (1.0 $<$ F090W$-$F150W $<$ 1.7~mag). The left-hand side shows completeness as a function of input F090W magnitude, indicating whether or not the artificial star would pass all of our quality criteria (in both filters) and show up as a genuine detection on our CMD. We find that the completeness at the level of the observed TRGB ($m_{\rm TRGB} =$ 26.90~mag) is relatively high (80$\%$). The right-hand side shows the difference between output and input F090W stellar magnitudes for the same set of artificial stars. We find that the photometric bias at the level of the observed TRGB is negligible ($<$0.01~mag).

\end{document}